\documentclass[
 reprint,
 amssymb, amsmath,
 aip,jcp
]{revtex4-1}

\usepackage{comment}
\usepackage{dcolumn}
\usepackage[nolist]{acronym}

\usepackage[colorlinks=true,linkcolor=blue]{hyperref}%
\expandafter\ifx\csname package@font\endcsname\relax\else
 \expandafter\expandafter
 \expandafter\usepackage
 \expandafter\expandafter
 \expandafter{\csname package@font\endcsname}%
\fi
\usepackage{xspace}

\usepackage{graphicx}
\usepackage{amsmath}

\usepackage[normalem]{ulem}
\usepackage[utf8]{inputenc}

\usepackage{tikz}
\usepackage{multirow}
\usepackage{bbold}

\usetikzlibrary{positioning}

\newcommand{\ket}[1]{{\ensuremath{|#1\rangle}\xspace}}
\newcommand{\bra}[1]{{\ensuremath{\langle #1|}\xspace}}
\newcommand{\braket}[2]{{\bra{#1} #2 \rangle\xspace}}
\newcommand{\elemm}[3]{\bra{#1} #2 \ket{#3}}
\newcommand{\basis}[0]{\mathcal{B}}
\newcommand{\basisl}[0]{\basis^l}
\newcommand{\basisr}[0]{\basis^r}

\newcommand{\eib}[0]{\tilde{E}_i^\basis}
\newcommand{\ei}[0]{\tilde{E}_i}

\newcommand{\ban}[1]{\hat{b}_{#1}}
\newcommand{\bac}[1]{\hat{c}^\dagger_{#1}}

\newcommand{\phiiu}[0]{ \ket{{\Phi_i}[u]}}
\newcommand{\Chiiu}[0]{ \ket{{\Chi_i}[u]}}

\newcommand{\pphii}[0]{{\Phi_i}}
\newcommand{\cchii}[0]{{\Chi_i}}

\newcommand{\Chi}{X}

\newcommand{\Egs}[0]{{\color{black}\tilde{E}_{\text{gs}}}}
\newcommand{\Phigs}[0]{{\color{black}\Phi_{\text{gs}}}}

\newcommand{\br}[1]{{\mathbf{r}_{#1}}}
\newcommand{\brbig}[1]{{\mathbf{R}_{#1}}}

\newcommand{\brb}[1]{{\bf r}_{#1}}
\newcommand{\rab}[1]{|\brb{1} - \brb{2}|}

\newcommand{\ai}[1]{\hat{a}_{#1}}
\newcommand{\aic}[1]{\hat{a}^{\dagger}_{#1}}

\newcommand{\nuclnum}{N_{\text{nucl}}}
\newcommand{\ku}[2]{\hat{K}[u](\br{#1},\br{#2})}

\newcommand{\lu}[3]{\hat{L}[u](\br{#1},\br{#2},\br{#3})}

\newcommand{\uu}[2]{u(\br{#1},\br{#2})}
\newcommand{\umu}[2]{u(r_{#1#2},\mu)}
\newcommand{\unew}[2]{\mathcal{U}(\br{#1},\br{#2},\mu,\{\alpha_i\})}
\newcommand{\envlop}[1]{\bar{g}(\br{#1})}
\newcommand{\tu}{\hat{\tau}_u}

\newcommand{\deriv}[3]{\frac{\partial^{#3} #1}{\partial {#2}^{#3}}}

\newcommand{\wmuijkl}[4]{{w}_{ij}^{kl}}
\newcommand{\vijkl}[0]{{V}_{ij}^{kl}}
\newcommand{\kijkl}[0]{{K}_{ij}^{kl}}

\newcommand{\lmuijmkln}[0]{{L}_{ijm}^{kln}}

\newcommand{\htc}[0]{\tilde{H}}
\newcommand{\htcz}[0]{\tilde{H}_0}
\newcommand{\vtc}[0]{\tilde{V}}
\newcommand{\hu}[0]{\tilde{H}[u]}
\newcommand{\hdagu}[0]{\tilde{H}^\dagger[u]}

\newcommand{\htcb}[0]{\tilde{H}^\basis}

\newcommand{\fderiv}[2]{\frac{\delta #1}{\delta{#2}}}

\newcommand{\etchiphi}[0]{\tilde{E}[\Chi,\Phi]}
\newcommand{\etchiphigs}[0]{\tilde{E}[\Chi,\Phigs]}

\newcommand{\etchi}[1]{\tilde{E}[\Chi,#1]}
\newcommand{\etchiz}[1]{\tilde{E}[\Chi_0,#1]}
\newcommand{\etphi}[1]{\tilde{E}[#1,\Phi]}

\newcommand{\rmi}[0]{{I}}

\newcommand{\tccipsi}[0]{E_{\rm{TC-CIPSI}}}
\newcommand{\tcfci}[0]{E_{\rm{TC-FCI}}}

\newcommand{\fci}[0]{E_{\rm{FCI}}}

\usepackage{csquotes}

\begin{document}

\author{Abdallah Ammar}
\email{aammar@irsamc.ups-tlse.fr}
\newcommand{\LCPQ}{Laboratoire de Chimie et Physique Quantiques (UMR 5626), Universit\'e de Toulouse, CNRS, UPS, France}
\affiliation{\LCPQ}
\author{Anthony Scemama}
\affiliation{\LCPQ}
\author{Emmanuel Giner}%
\email{emmanuel.giner@lct.jussieu.fr}
\affiliation{Laboratoire de Chimie Théorique, Sorbonne Université and CNRS, F-75005 Paris, France}

\newcommand{\Ort}{orthogonal}
\newcommand{\BiO}{bi-orthogonal}

\title{Transcorrelated selected configuration interaction in a bi-orthonormal basis and a cheap three-body correlation factor}


\begin{abstract}
In this work, we develop a mathematical framework for a Selected Configuration
Interaction (SCI) algorithm within a \BiO{} basis for
transcorrelated (TC) calculations. The \BiO{} basis used here serves as the
equivalent of the standard Hartree Fock (HF) orbitals. However, within the
context of TC, it leads to distinct orbitals for the left and right vectors.
Our findings indicate that the use of such a \BiO{} basis allows for a proper
definition of the frozen core approximation. In contrast, the use of HF
orbitals results in bad error cancellations for ionization potentials
and atomization energies (AE). Compared to HF
orbitals, the optimized \BiO{} basis significantly reduces the positive part of
the second-order energy (PT2), thereby facilitating the use of standard
extrapolation techniques of hermitian SCI.
While we did not observe a significant improvement in the convergence of the
SCI algorithm, this is largely due to the use in the present work of a simple three-body
correlation factor introduced in a recent study. This correlation factor, which
depends only on atomic parameters, eliminates the need for re-optimization of
the correlation factor for molecular systems, making its use straightforward
and user-friendly.
Despite the simplicity of this correlation factor, we were able to achieve
accurate results on the AE of a series of 14 molecules in a triple-zeta basis.
We also successfully broke a double bond until the full dissociation limit
while maintaining the size consistency property. This work thus demonstrates
the potential of the BiO-TC-SCI approach in handling complex molecular systems.
\end{abstract}

\maketitle

\begin{acronym}
  \acro{FCI}{Full Configuration Interaction}
  \acro{FROGG}{frozen Gaussian geminal}
  \acro{CI}{Configuration Interaction}
  \acro{SCI}{Selected \ac{CI}}
  \acro{EN}{Epstein Nesbet}
  \acro{QMC}{Quantum Monte Carlo}
  \acro{AO}{Atomic Orbital}
  \acro{MO}{Molecular Orbital}
  \acro{HF}{Hartree-Fock}
  \acro{CAS}{Complete Active Space}
  \acro{VMC}{Variational Monte Carlo}
  \acro{DMC}{Diffusion Monte Carlo}
  \acro{TC}{Transcorrelated}
  \acro{CASSCF}{complete active space self consistent field}
  \acro{fc}{frozen core}
  \acro{FCIQMC}{Full Configuration Interaction Quantum Monte Carlo}
  \acro{SCF}{Self Consistent Field}
  \acro{RHF}{Restricted   Hartree-Fock}
  \acro{UHF}{Unrestricted Hartree-Fock}
  \acro{ROHF}{Restricted Open-shell Hartree-Fock}
  \acro{DIIS}{Direct Inversion in the Iterative Subspace}
  \acro{LS}{Level-Shifting}
  \acro{MP}{M{\o}ller-Plesset}
  \acro{MPPT2}[MP2]{M{\o}ller-Plesset Perturbation Theory at second order}
  \acro{BiO}{Bi-Orthogonal}
  \acro{Ort}{Orthogonal}
  \acro{MPS}{Matrix Product State}
  \acro{MBPT2}[MBPT(2)]{Many Body Perturbation Theory to the second order}
  \acro{LCCSD}{Linearized Coupled Cluster Singles and Doubles}
  \acro{DMRG}{Density Matrix Renormalization Group}
  \acro{CC}{Coupled Cluster}
  \acro{DFT}{Density Functional Theory}
  \acro{WFT}{Wave Function Theory}
  \acro{Var}{Variational}
  \acro{CBS}{Complete Basis Set}
  \acro{IP}{Ionization Potential}
  \acro{AE}{Atomization Energy}
\end{acronym}

\section{Introduction}

In the expansive field of electronic structure calculations,
the \ac{TC} method~\onlinecite{Boys_Handy_1969_determination} offers an appealing path toward an accurate description
of atomic and molecular systems.
The \ac{TC} framework uses
a linear combination of Slater determinants multiplied by a correlation factor, thus enabling the inclusion of correlation effects
with both real space space and orbital space representations.
The specificity of the \ac{TC} methodology is its explicit incorporation of the correlation factor's
effects into the Hamiltonian via a similarity transformation~\onlinecite{Hirschfelder-JCP-63, Boys_Handy_1969_determination}.
This yields a non-Hermitian \ac{TC} Hamiltonian, which although forbids the use of the variational principle
for wave function optimization, grants the capacity to make the wave function's expression more compact,
and accelerates the convergence toward the complete basis set limit.
Furthermore, the effective interaction within the \ac{TC} 
Hamiltonian is limited to three-body terms at most, which facilitates the
deterministic calculation of all integrals necessary for the optimization of
the wave function. This characteristic eliminates the requirement for
stochastic sampling of the $N$-body integrals, a process typically seen in
\ac{VMC} methods.

There are two distinct but connected aspects involved in the optimization of the wave function
within the \ac{TC} framework:
i) the optimization of the correlation factor used to perform the similarity transformation
and ii) the optimization of the Slater-determinant part of the wave function.

The role of the correlation factor is quite simple to understand:
it fundamentally involves decreasing the probability of finding two electrons at small inter-electronic distances.
However, there exists a wide variety of functional forms for the correlation factor which can be
broadly divided into two categories: universal two-body
correlation factors and three-body correlation factors. Universal correlation factors dependent solely
on inter-electronic coordinates and create an homogeneous and isotropic correlation hole throughout the whole space.
In contrast, three-body correlation factors present a richer parametrisation,
enabling for instance the adaptation of the correlation hole's depth and extension based on
the distance of an electron pair from a specific nucleus.

The \ac{FROGG}, proposed by Ten-No~\onlinecite{TenNo_2000_feasible}, was optimized for valence electron pairs and
can be seen as the prototype for the universal correlation factor. An alternative universal correlation factor
was derived from the range-separated density functional theory~\onlinecite{Gin-JCP-21}, which depends on a
single parameter $\mu$ that tailors the shape of the correlation hole.
The utilization of a universal correlation factor is appealing due to its
minimal optimization requirements. However, it is crucial to note that when
enforcing a correlation hole designed for valence electrons in high-density
regions, there is a significant demand for flexibility in the wave function,
enough to adapt the density in the core regions.
Conversely, three-body correlation factors (for instance,
see Refs.~\onlinecite{Boys_Handy_1969_calculation,MosSchLeeKal-JCP-82-Monte,SchMos-JCP-90,UmrWilWil-PRL-88,MosSch-JCP-92,MusNig-JCP-94,HuaUmrNig-JCP-97,GucJeoUmrJai-APS-05,LopSetDruNee-APS-12,AusZubLes-CR-12,LucStuSchHag-JCP-2015}
and references therein)
employ explicit electron-electron-nucleus coordinates, allowing for the
adjustment of the correlation hole's depth and spatial extension according to
the system's density, albeit at the expense of extensive optimization.
Recently, we introduced a minimal three-body correlation factor~\cite{AmmSceGin-JCTC-2023}
which necessitates limited optimization. The performance of this correlation factor
will be investigated here.

When it comes to optimizing the Slater-determinant part of the \ac{TC}
wave function, it is important to note that the \ac{TC} Hamiltonian, as any $N$-body operator,  can be
expressed in second quantization. The latter implies that any conventional wave
function ansatz from quantum chemistry can be adapted to fit within the \ac{TC}
framework.
However, the bi-orthonormal nature of the left and right eigenvectors
in the \ac{TC} Hamiltonian suggests potential change of strategy to choose the basis.
Specifically, the latter could be a pair of bi-orthonormal one-electron
basis sets to expand both the left and right eigenvectors. This is in contrast to the
standard practice of using a single orthonormal one-electron basis set.

Furthermore, replacing a single set of orbitals with two distinct and
optimizable sets increases the flexibility of both left and right
eigenvectors. This change leads to a decrease in the error associated with the
\ac{TC} energy, which is directly related to the product of the
errors in the left and right eigenvectors. 
Constraining these two sets to form a couple of bi-orthonormal basis helps 
in defining creation and annihilation operators. These operators fulfill the
anti-commutation rules of standard fermionic operators in second-quantized
formulations, simplifying the process of adapting any conventional wave
function method to the \ac{TC} framework. 

Another important aspect of the \ac{TC} framework is that the variational principle does not hold anymore 
due to the non hermitian nature of the \ac{TC} Hamiltonian. One can nevertheless substitute the energy minimization 
by the so-called bi variational functional~\cite{Boys_Handy_1969_determination}. 
The latter involves finding stationary points of an energy functional depending on both a left and right function. 
As an example, the optimization of the energy with respect to orbital parameters at the single determinant level 
necessary leads to two sets of bi orthonormal orbitals adapted for either the left or right eigenvectors. 

Taking these peculiar aspects into consideration, we distinguish here between formalisms that use left and right orbitals 
(termed \emph{\BiO{}} frameworks) and those that employ a common set of orbitals (denoted as \emph{\Ort{}} formalisms).

Historically, an \Ort{} framework utilizing a single Slater determinant was often employed
\cite{Boys_Handy_1969_determination,Boys_Handy_1969_calculation,Boys_Handy_1969_condition,
Boys_Handy_1969_first,Handy_1969,Boys_1969,Handy_1971_minimization,Handy_1972,Armour_1972_application,Bernardi_Boys_1973,Bernardi_1973,Armour_1974,Handy_1975,Huggett_Armour_1976,DharmaWardana_1976,Careless_etal_1977,Hall_Miller_1978}.
Fimple and Unwin's pioneering work~\cite{Fimple_Unwin_1976} introduced a \ac{CI} expansion
within a \BiO{} framework, illustrated on the simple case of the ground state of the Helium atom.
Subsequently, the \BiO{} formalism was significantly expanded by Ten-No \textit{et al} with
\ac{MPPT2} and \ac{LCCSD} using the \ac{FROGG} correlation factor in the early 2000s~
\cite{TenNo_2000_feasible,Hino_etal_2001,Hino_etal_2002,TenNo_Hino_2002}. These advances,
coupled with efforts on the three-body terms~\cite{TenNo_2000_three}, enabled calculations
on small organic molecules. Recent developments using an \Ort{} framework were proposed by several groups
using \ac{FCIQMC}~\cite{Cohen_etal_2019,Guther_etal_2021,DobCohAlaGin-JCP-22},
\ac{CC}~\cite{Schraivogel_etal_2021,Schraivogel_etal_2023},
\ac{MPS}~\cite{BaiRei-JCP-20-Transcorrelated,BaiLesRei-JCTC-22,LiaZhaChrSchRioKatAla-JCTC-23} or \ac{CI}~\cite{AmmGinSce-JCTC-22}.
Periodic systems also saw applications using either a single Slater determinant, \ac{MPPT2}
, \ac{CI} or \ac{CC} within the \Ort{} framework~
\cite{Armour_1980,Umezawa_Tsuneyuki_2004_ground,Sakuma_Tsuneyuki_2006,Tsuneyuki_2008,Ochi_etal_2012,Luo_2012,
Ochi_Tsuneyuki_2013,Ochi_etal_2014,Ochi_Tsuneyuki_2014,WahlenStrothman_etal_2015,Ochi_Tsuneyuki_2015,Ochi_etal_2016,
Jeszenszki_etal_2018,Luo_Alavi_2018,Dobrautz_etal_2019,Jeszenszki_etal_2020,Liao_etal_2021},
and later developments using a \BiO{} framework were proposed
\cite{Ochi_etal_2017,Ochi-CPC-23}.

In one of our previous work~\cite{AmmSceGin-JCP-22}, we introduced
\ac{SCI} for the \ac{TC} framework using a single set of orthonormal molecular orbitals within a bi-variational scheme.
We observed that enhancing the quality of both the left and right wave
functions stabilizes the computation of the second-order perturbative
correction. The latter is crucial as it makes the \ac{SCI} approach competitive
by significantly improving the convergence rate of calculations and enabling
extrapolation techniques~\cite{HolUmrSha-JCP-17}.
The quality of the \acp{MO} is a critical aspect for the practical application
of \ac{SCI}.  Specifically, the optimization of orbitals in \ac{SCI} allows for
a more compact \ac{CI} expansion, either through variationally optimized
orbitals~\cite{SmiMusHolSha-JCTC-17,YaoUmr-JCTC-21,Par-JCTC-21,GuoZhaLeiLiu-JCTC-21}
or state-average natural orbitals for excited
state excitation energies~\cite{LooSceBloGarCafJac-JCTC-18,LooLipBogSceJac-JCTC-20}.
We would also like to stress the importance of orbital
optimization, especially in the context of one of \ac{SCI}'s primary
applications: achieving near CASSCF with a large active
space~\cite{SmiMusHolSha-JCTC-17,Par-JCTC-21,GuoZhaLeiLiu-JCTC-21}.
When considering orbital optimization in a bi-variational framework,
the \BiO{} framework is a natural choice. Hence, the present article aims to
further develop \ac{SCI} within a fully bi-orthonormal and bi-variational
framework, i.e. with different left and right molecular orbitals.

In this work, we use a linearized version of the recently developed cheap
three-body correlation factor~\cite{AmmSceGin-JCTC-2023}. However, we want to
highlight that the \ac{BiO}-\ac{SCI} proposed here can also be applied with more
sophisticated correlation factors, leading to improved convergence properties
of the present method.

The structure of the article is the following.
Section \ref{sec:theo_tc} provides a brief overview of the \ac{TC} framework, then Section \ref{subsubsec:stationary}
gives a summary of the bi-variational principle, which is formalized in \BiO{} framework in Section \ref{sec:bio}
and then applied to the \ac{SCI} algorithm in Section \ref{sec:bio_tc_cipsi}. We give the explicit form of the correlation
factor used here in Section \ref{sec:three_e_j}.
We present the results in Section \ref{sec:results}, which begins with an investigation of the benefit of using
a \BiO{} framework for \ac{SCI} in Section \ref{sec:benefit_bio} specially regarding the \ac{fc} approximation
which is fundamental in view to apply the \ac{TC} framework to large systems.
Then, we illustrate the convergence of the present BiO-TC-SCI algorithm in Section \ref{sec:conv_tc_cipsi}
on the F$_2$, N$_2$ and CO molecules, and show that the usual linear extrapolation is possible within our framework.
We continue our study by computing the dissociation curve of the CO molecule, which allows us to test the present
framework in different regimes of correlation together with the size consistency property.
We conclude the present work by computing the atomization energies of a set of 14 molecules in increasing basis sets
and compare with both the usual \ac{SCI} algorithm and estimated exact results. We observe that in most of the case,
the accuracy of a calculation in the cc-pVTZ basis set is within 1 kcal/mol with respect to the estimated exact results.
We emphasize that no optimization of the correlation factor was needed for molecular systems thanks
to the specific form of the correlation factor used here~\cite{AmmSceGin-JCTC-2023}.
Eventually, we summarize the main results in Section \ref{sec:conclu}.

\section{Theory}
\label{sec:theo}
\subsection{Basics of the transcorrelated formalism}
\label{sec:theo_tc}
The general form of the transcorrelated Hamiltonian for a symmetric correlation factor $\uu{1}{2}$ is given by
\begin{equation}
 \label{ht_def_g}
 \begin{aligned}
  \hu  &\equiv e^{-\tu} \hat{H} e^{\tu} \\
                & = \hat{H} + \big[ \hat{H},\tu \big] + \frac{1}{2}\bigg[ \big[\hat{H},\tu\big],\tu\bigg],
 \end{aligned}
\end{equation}
where $\tu = \sum_{i<j}u(\br{i},\br{j})$ and $\hat{H} = \sum_i -\frac{1}{2} \nabla^2_i + v(\br{}_i) + \sum_{i<j}   1/r_{ij}$.
Eq. \eqref{ht_def_g} leads to the following transcorrelated Hamiltonian
\begin{equation}
 \begin{aligned}
 \label{ht_def_g2}
 \hu& = \hat{H} - \sum_{i<j} \ku{i}{j} - \sum_{i<j<k} \lu{i}{j}{k},
 \end{aligned}
\end{equation}
where the effective two- and three-body operators $\ku{1}{2}$ and $\lu{1}{2}{3}$ are defined as
\begin{equation}
 \label{def:k_u}
 \begin{aligned}
  \ku{1}{2} = \frac{1}{2} \bigg( &\Delta_1 \uu{1}{2} + \Delta_2 \uu{1}{2} \\
                                         + &\big(\nabla_1 \uu{1}{2} \big) ^2 + \big(\nabla_2 \uu{1}{2}      \big) ^2 \bigg) \\
                                         + &\nabla_1 \uu{1}{2} \cdot \nabla_1 + \nabla_2 \uu{1}{2}\cdot     \nabla_2,
 \end{aligned}
\end{equation}
and
\begin{equation}
 \label{def:l_u}
 \begin{aligned}
  \lu{1}{2}{3} = &   \nabla_1 \uu{1}{2} \cdot \nabla_1 \uu{1}{3} \\
                                          + & \nabla_2 \uu{2}{1} \cdot \nabla_2 \uu{2}{3}   \\
                                          + & \nabla_3 \uu{3}{1} \cdot \nabla_3 \uu{3}{2}   .
 \end{aligned}
\end{equation}
As apparent from the definition of Eq.\eqref{ht_def_g}, $\hu $ is not Hermitian as
\begin{equation}
 \hdagu = e^{+\tu} \hat{H} e^{-\tu} \ne \hu,
\end{equation}
and a given eigenvalue $\ei$ is associated with a couple of right and left eigenvectors
\begin{equation}
 \label{eq:tc_eigv}
 \begin{aligned}
 & \hu \phiiu = \ei \phiiu \\
 & \hdagu \Chiiu = \ei \Chiiu.\\
 \end{aligned}
\end{equation}
Nevertheless, thanks to the property of similarity transformations, the spectrum of $\hu$ coincides with that of the usual Hamiltonian.
From thereon, we no longer include the explicit dependence on the correlation factor $u$, and instead, we will use $\htc$, $\ket{\Chi_i}$, and $\ket{\Phi_i}$ to represent the TC Hamiltonian and its corresponding left and right eigenvectors.

\subsection{Non-Hermitian eigenvalue problems and the bi-variational principle }
\label{subsubsec:stationary}
Due to the non-Hermitian nature of the \ac{TC} Hamiltonian, the standard energy minimization approach cannot be
used to optimize wavefunctions since the variational principle does not apply.
Instead, the search for an energy minimum over a wavefunction $\Psi$ can be replaced by the search for
a stationary point of a functional $\etchiphi$ that depends on two wavefunctions $\Chi$ and $\Phi$
\begin{equation}
 \label{e_phichi}
 \etchiphi = \frac{\elemm{\Chi}{\htc}{\Phi}}{\braket{\Chi}{\Phi}},
\end{equation}
and $\Chi$ and $\Phi$ are often referred to as the left and right wavefunctions, respectively.
An eigenvalue $\ei$ is obtained when either the left or right function is an eigenfunction.
\begin{equation}
 \begin{aligned}
 \etchi{\pphii} = \frac{\elemm{\Chi}{\htc}{\pphii}}{\braket{\Chi}{\pphii}} = \ei\,\, \forall \Chi, \\
 \etphi{\cchii} = \frac{\elemm{\Chi_i}{\htc}{\Phi}}{\braket{\Chi_i}{\Phi}} = \ei\,\, \forall \Phi. \\
 \end{aligned}
\end{equation}
Hence, finding $\ei$ and the corresponding left and right eigenvectors is equivalent to nullifying the right and left functional derivatives
\begin{equation}
 \begin{aligned}
   \fderiv{\etchi{\pphii}}{\Chi} = 0 \,\, \forall  \,\, \Chi, \\
   \fderiv{\etphi{\cchii}}{\Phi} = 0 \,\,  \forall  \,\, \Phi, \\
 \end{aligned}
\end{equation}
which have general forms given by
\begin{equation}
 \label{deriv_chi}
 \fderiv{\etchiphi}{\Chi} =
 \frac{\left( \htc \Phi \right) \, \braket{\Chi}{\Phi} - \elemm{\Chi}{\htc}{\Phi} \Phi}
      {|\braket{\Chi}{\Phi}|^2},
\end{equation}
\begin{equation}
 \label{deriv_phi}
 \fderiv{\etchiphi}{\Phi} =
 \frac{\left( \htc^\dagger \Chi \right) \, \braket{\Chi}{\Phi} - \elemm{\Chi}{\htc}{\Phi} \Chi}
      {|\braket{\Chi}{\Phi}|^2}.
\end{equation}
This is the so-called bi-variational principle~\cite{Boys_Handy_1969_determination}.
It is worth highlighting that setting the \emph{left} functional derivative to zero \emph{for all left wavefunctions $\Chi$} enables the determination of the \emph{optimal right wavefunction} (and \textit{vice versa}).

\subsection{Transcorrelation in a bi-orthonormal framework}
\label{sec:bio}
In this section we briefly summarize how the \ac{TC} Hamiltonian is written in a bi-orthonormal basis
(see Sec.~\ref{sec:lef-right_det}) and also recall the mathematical framework together with the 
self-consistent field equations used to obtain the bi-orthonormal basis (see Sec.~\ref{sec:orb_opt}).
\subsubsection{Bi-orthonormal framework and second quantization}
\label{sec:lef-right_det}
In practice, the TC Hamiltonian is projected into a one-particle basis set $\basis$
\begin{equation}
 \label{eq:def_hub}
 \begin{aligned}
 &\htcb   = \hat{P}^{\basis} \htc \hat{P}^\basis,
 \end{aligned}
\end{equation}
where $\hat{P}^{\basis}$ is the projector onto the Hilbert space spanned by the one-particle basis set $\basis$.
Because of the properties of the similarity transformation the exact eigenvalue $E_i$ is recovered in the \ac{CBS} limit
\begin{equation}
 \lim_{\basis \rightarrow \text{CBS}} \eib = E_i,
\end{equation}
and given that some of the correlation effects are accounted for by the correlation factor, we can anticipate a faster convergence of $\eib$ compared to wave function methods based on the standard Hamiltonian. For the sake of simplicity in notation, we will henceforth drop the exponent $\basis$ and refer directly to $\htcb$ as $\htc$.

In a second quantization framework, rather than using a standard basis set consisting of real-valued orthonormal spatial \acp{MO} ${\phi_i(\br{})}$ to express the $\htc$ operator, a more flexible approach involves using distinct \acp{MO} for the left and right functions.
The left wave functions are expanded on the set of \emph{left} real-valued orbitals $\basisl=\{\chi_i(\br{}),i=1,\dots,n \}$
while the right wave functions are expanded on the set of \emph{right} real-valued orbitals $\basisr=\{ \phi_i(\br{}),i=1,\dots,n \}$.
Similar to the usual orthonormal framework, if the two bases $\basisl$ and $\basisr$ are chosen to satisfy the biorthonormal relation $\braket{\chi_i}{\phi_j} = \delta_{ij}$,
one can build creation operators $\bac{k,\sigma}$ and annihilation operators $\ban{l,\lambda}$
(where $k$ and $l$ are labels of spin-free orbitals, and $\sigma$ and $\lambda$ are labels of spins) that satisfy the usual 
anticommutation relations~\cite{Moshinsky_Seligman_1971,Gouyet_1973_champ,Fimple_Unwin_1976,Dahl_1978,Payne_1982,Hino_etal_2001,Surjan_2011}
\begin{equation}
\begin{aligned}
        \label{eq:commut_b}
        [\bac{k,\sigma},\bac{l,\lambda}]_{+} &= 0, \\
        [\ban{k,\sigma},\ban{l,\lambda}]_{+} &= 0, \\
        [\bac{k,\sigma},\ban{l,\lambda}]_{+} &= \delta_{kl} \delta_{\sigma \lambda},
\end{aligned}
\end{equation}
and avoid the complications introduced by overlap integrals inherent to the use of
non-orthonormal basis
functions~\cite{Takano_1959,Cantu_etal_1975,Kvasnicka_1977,Kojo_Hirose_2009}.
As a result, expressing an operator in second quantization using the biorthonormal bases $\basisl$ and $\basisr$ involves two straightforward rules: (i) replace the conventional creation and annihilation operators $\aic{k,\sigma}$ and $\ai{k,\sigma}$ with the biorthonormal basis creation and annihilation operators $\bac{k,\sigma}$ and $\ban{k,\sigma}$, and (ii) write the integrals of the operator using the functions ${\chi_i}$ in the bra and ${\phi_j}$ in the ket.

Following these rules the $\htc$ operator can be written in a second-quantized form on the two bi-orthonormal basis sets as follows
\begin{equation}
 \label{eq:def_hub_sec_q}
 \begin{aligned}
 &\htc  = \sum_{k \in \basisl} \sum_{i \in \basisr} \sum_{\sigma \in \{ \uparrow,\downarrow \}} h_{ki}\, \bac{k,\sigma}\,\ban{i,\sigma}\\
 & + \frac{1}{2}\sum_{k,l \in \basisl} \sum_{i,j \in \basisr} \sum_{\sigma,\lambda \in \{ \uparrow,\downarrow \}}
 \big( V_{ij}^{kl} - \kijkl\big)\, \bac{k,\sigma}\, \bac{l,\lambda}\, \ban{j,\lambda}\, \ban{i,\sigma} \\
       & - \frac{1}{6} \sum_{k,l,n \in \basisl} \sum_{i,j,m \in \basisr} \sum_{\sigma,\lambda,\kappa \in \{ \uparrow,\downarrow \}}
\lmuijmkln\, \bac{k,\sigma}\, \bac{l,\lambda}\, \bac{n,\kappa}\, \ban{m,\kappa}\, \ban{j,\lambda}\, \ban{i,\sigma},
 \end{aligned}
\end{equation}
where $h_{ki}$ and $V_{ij}^{kl}$ are the integrals of the usual one- and two-electron operators, respectively, expressed in the bi-orthonormal basis,
\begin{equation}
 h_{ki} = \int \text{d} \br{}\, \chi_k(\br{ })\, \hat{h}\, \phi_i(\br{ }),
\end{equation}
\begin{equation}
 \vijkl = \int \text{d} \br{1}\, \text{d} \br{2}\, \chi_k(\br{1})\, \chi_l(\br{2})\, \frac{1}{r_{12}}\, \phi_i(\br{1})\, \phi_j(\br{2}),
\end{equation}
and $\kijkl$ and $\lmuijmkln$ are the two- and three-electron integrals corresponding to the effective two- and three-body operators $\ku{1}{2}$
and $\lu{1}{2}{3}$ expressed in the bi-orthonormal basis
\begin{equation}
 \kijkl = \int \text{d} \br{1}\, \text{d} \br{2}\, \chi_k(\br{1})\, \chi_l(\br{2})\, \ku{1}{2}\, \phi_i(\br{1})\, \phi_j(\br{2}),
\end{equation}
\begin{equation}
 \begin{aligned}
  \lmuijmkln  = \int \text{d} \br{1}\, \text{d} \br{2}\, \text{d} & \br{3}\, \chi_k(\br{1})\, \chi_l(\br{2})\, \chi_n(\br{3})\, \\ & \lu{1}{2}{3}\, \phi_i(\br{1})\, \phi_j(\br{2})\, \phi_m(\br{3}).
 \end{aligned}
\end{equation}

An important consequence of the bi-orthonormal framework is the one-to-one correspondence between
the left  and right Slater determinants even if they are built with different functions.
To illustrate this, let us consider a left Slater determinant
$\bra{\Chi_I} = \prod_{i\in \mathcal{S}_{\Chi_I} }^N \bra{0}\ban{i} $
where $\mathcal{S}_{\Chi_I} = \{ k \}$ is the ordered list of indices of left orbitals $\chi_k(\br{})$ occupied in $\bra{\Chi_I}$.
In a similar way, let us consider a right Slater determinant
$\ket{\Phi_J} = \prod_{i\in \mathcal{S}_{\Phi_J} }^N \bac{i} \ket{0}$
where $\mathcal{S}_{\Phi_J} = \{ l \}$ is the ordered list of indices of right orbitals $\phi_l(\br{})$ occupied in $\ket{\Phi_J}$.
Because of the bi-orthonormality relation, these two Slater determinants are orthogonal only if the two sets
$\mathcal{S}_{\Chi_I}$ and $\mathcal{S}_{\Phi_J}$ are different
\begin{equation}
 \begin{aligned}
 \braket{\Chi_I}{\Phi_J} & = 0 \quad \text{if } \mathcal{S}_{\Chi_I} \ne \mathcal{S}_{\Phi_J} , \\
                         & = 1 \quad \text{if } \mathcal{S}_{\Chi_I} = \mathcal{S}_{\Phi_J} .\\
 \end{aligned}
\end{equation}
As a consequence, to each right Slater determinant $\ket{\Phi_I}$ one can associate a unique
left Slater determinant $\ket{\Chi_I}$ with an identical list of indices of occupied orbitals, although
the orbitals composing these two Slater determinants are themselves different.

\subsubsection{Optimization of bi-orthonormal orbitals in the TC framework}
\label{sec:orb_opt}
The selection of orbitals used to expand the wave function is a crucial factor
in enhancing the accuracy of any approximated wave function ansatz. In the
Hermitian case, the starting point is often the set of \ac{HF}
orbitals, as they minimize the energy of a single Slater determinant wave
function.
However, in the case of the TC Hamiltonian, which is a non-Hermitian operator,
the expectation value of a single Slater determinant cannot be minimized.
Instead, one can seek a stationary point of the functional $\etchiphi$, where
both $\Chi$ and $\Phi$ are single-determinant wave functions. 
It's important to note that these wave functions are not necessarily built with the 
same orbitals. 
Using reference determinants $\Chi_0$ and $\Phi_0$ constructed on
two bi-orthonormal basis sets, $\basisl$ and $\basisr$ respectively, and with
the same orbital occupancy,
we can write the left and right single Slater determinant wave function as follows
\begin{equation}
 \begin{aligned}
 &\ket{\Phi[\hat{\kappa}]} = e^{\hat{\kappa}} \ket{\Phi_0},\\
 &\ket{\Chi[\hat{\kappa}]} = e^{\hat{\kappa}}\ket{\Chi_0}  .\\
 \end{aligned}
\end{equation}
Here $\hat{\kappa}$ are anti hermitian orbital rotation operators  
\begin{equation}
 \hat{\kappa} = \sum_{p>q} \kappa_{pq} \hat{{E}}_{pq},
\end{equation}
with the operators $\hat{{E}}_{pq}$ being defined as 
\begin{equation}
\begin{aligned}
 & \hat{{E}}_{pq}  =  \sum_{\sigma \in \{ \uparrow,\downarrow \}} \bac{p,\sigma}\ban{q,\sigma}, \\ 
\end{aligned}
\end{equation}
and where the $\kappa_{pq}$ are the orbital rotation parameters forming the matrix $\boldsymbol{\kappa}$. 
The fact that we use a single set of orbital parameters $\boldsymbol{\kappa}$ even if we optimize two distinct functions 
is due to the bi-orthonormal condition which ensures 
\begin{equation}
 \label{eq:bi_ort_sd}
 \begin{aligned}
\braket{\Chi[\hat{\kappa}]}{\Phi[\hat{\kappa}]} & = \elemm{\Chi_0}{e^{-\hat{\kappa}}e^{\hat{\kappa}}}{\Phi_0} \\ 
             & =\braket{\Chi_0}{\Phi_0}   =1. 
 \end{aligned}
\end{equation}
In other words, the bi orthonormal condition imposes that the left orbitals 
are related to the right orbitals through the matrix $\mathbf{S}^{-1}$, where $\mathbf{S}$ is the overlap 
matrix between the right orbitals. 
Another point is that the matrix of the orbital parameters $\boldsymbol{\kappa}$ is no longer anti hermitian but fulfills 
a pseudo anti hermitian relation 
\begin{equation}
 \big(\boldsymbol{\kappa}\big)^\dagger = - \mathbf{S} \boldsymbol{\kappa} \mathbf{S}^{-1}.
\end{equation}

The equivalent of the standard energy minimization condition in the \ac{HF} framework 
translates within the \ac{TC} context in seeking a stationary point of the following functional:
\begin{equation}
 \tilde{E}\big[\hat{\kappa}\big] =
\frac{\elemm{\Chi_0}{e^{-\hat{\kappa}}\htc e^{\hat{\kappa}}}{\Phi_0}}
{ \elemm{\Chi_0}{e^{-\hat{\kappa}}e^{\hat{\kappa}}}{\Phi_0} }.
\end{equation}
and, using Eq. \eqref{eq:bi_ort_sd}, 
the functional $\tilde{E}\big[\hat{\kappa}\big]$ can be written up to first-order in $\hat{\kappa}$ as
\begin{equation}
 \label{eq:first_order}
\tilde{E}\big[\hat{\kappa}\big] =  
\elemm{\Chi_0}{\htc }{\Phi_0} -  \elemm{\Chi_0}{\hat{\kappa}\htc }{\Phi_0} + \elemm{\Chi_0}{\htc \hat{\kappa}}{\Phi_0} 
+ o(|\hat{\kappa}|^2). 
\end{equation}
Differentiating Eq. \eqref{eq:first_order} with respect to $\kappa_{pq}$ and evaluating at $\boldsymbol{\kappa}=0$ 
leads to
\begin{equation}
 \label{eq:derivative}
 \deriv{\tilde{E}\big[\hat{\kappa}\big]}{\kappa_{pq}}{}\bigg|_{\boldsymbol{\kappa}=0} 
   = \elemm{\Chi_0}{\htc\hat{{E}}_{pq}}{\Phi_0} - \elemm{\Chi_0}{\hat{{E}}_{pq}\htc}{\Phi_0}.
\end{equation}
We assume here closed shell determinants and label the occupied and virtual orbitals by $i$ and $a$, respectively. 
By noticing that $\bra{\Chi_0}\hat{{E}}_{ai}= 0 = \hat{{E}}_{ia} \ket{\Phi_0} $,  
canceling the derivative with respect to occupied-virtual orbital rotation parameters leads to 
\begin{equation}
 \begin{aligned}
 & \deriv{\tilde{E}\big[\hat{\kappa}\big]}{\kappa_{ia}}{}\bigg|_{\boldsymbol{\kappa}=0} = 0 
\Leftrightarrow & \elemm{\Chi_0}{\hat{{E}}_{ia}\htc}{\Phi_0} = 0, 
 \end{aligned}
\end{equation}
and 
\begin{equation}
 \begin{aligned}
 & \deriv{\tilde{E}\big[\hat{\kappa}\big]}{\kappa_{ai}}{}\bigg|_{\boldsymbol{\kappa}=0} = 0  
\Leftrightarrow & \elemm{\Chi_0}{\htc\hat{{E}}_{ai}}{\Phi_0} = 0, 
 \end{aligned}
\end{equation}
which give the left and right Brillouin conditions for the set of orbitals composing $\ket{\Chi_0}$ and $\ket{\Phi_0}$.  
Of course, all this derivation is trivially extendable to the open shell case. 

Just like in the Hermitian case, these Brillouin conditions can be fulfilled by
iteratively diagonalizing a Fock-like operator.  Within the \ac{TC} framework, the
corresponding Fock operator is non-Hermitian and constructed not with a standard
density, but with a transition density between the left and right functions.
Using matrix notations, the transition density $\mathbf{D}$ can be written in the \ac{AO} basis as
\begin{equation}
 \mathbf{D} = \mathbf{C}_L^\dagger \mathbf{C}_R,
\end{equation}
where $\mathbf{C}_L$ ($\mathbf{C}_R$) represents the matrix of coefficients of occupied left (right) orbitals on the \ac{AO} basis. 
The interested reader can look for instance in Ref.~\onlinecite{AmmSceGin-JCTC-2023} where an explicit form of the Fock operator is given.

\subsection{Development of selected CI in a bi-orthonormal framework}
\label{sec:bio_tc_cipsi}
In this section we describe the theoretical background of the TC-\ac{SCI} in the bi-orthonormal framework.
This requires first the development of a Rayleigh-Schrödinger perturbation theory using the bi-variational principle as the
starting point and adapting it to a bi-orthonormal basis (Sec.~\ref{sec:pt_bilin}).
We then give the explicit algorithm used for our \ac{BiO}-TC-\ac{SCI} algorithm in Sec.~\ref{sec:tcsci},
together with the technical details in Sec.~\ref{sec:technique}.

\subsubsection{Perturbation theory of the functional}
\label{sec:pt_bilin}
Following Ref.~\onlinecite{AmmSceGin-JCP-22} we give the perturbation expansion of the functional $\etchiphi$
for the ground-state energy. Here, the derivation is adapted to the bi-orthonormal framework.

Let the Hamiltonian $\htc$ be split into
\begin{equation}
 \htc = \htcz + \lambda \vtc.
\end{equation}
We assume that the left and right eigenvectors of $\htcz$ are known,
\begin{equation}
 \begin{aligned}
 &\htcz \ket{\Phi_i} = \epsilon_i \ket{\Phi_i} \\
 &\htcz^\dagger \ket{\Chi_i} = \epsilon_i \ket{\Chi_i}, \\
 \end{aligned}
\end{equation}
and that they form a bi-orthonormal set
\begin{equation}
 \braket{\Chi_i}{\Phi_j} = \delta_{ij}.
\end{equation}
In the case of a \ac{SCI} theory within an \ac{EN}\cite{epstein,nesbet} partitioning, $\ket{\Chi_0}$  and $\ket{\Phi_0}$ 
are the ground state left and right eigenvectors of $\htcz$ (which can be multi configurational) 
while the $\ket{\Chi_i}$ and $\ket{\Phi_i}$ for $i>0$ are Slater determinants 
such that $\braket{\Chi_0}{\Phi_i} = \braket{\Chi_i}{\Phi_0}=0$.   
The ground-state energy $\Egs$ can be obtained by evaluating $\etchiphigs$, \textit{i.e.} evaluating 
the functional at the right ground-state wave function, but the choice of $\Chi$ remains. 
Based on our previous numerical study\cite{AmmSceGin-JCP-22}, we choose $\ket{\Chi}=\ket{\Chi_0}$, \textit{i.e.} 
the left eigenfunction of $\htcz$. The functional then reads 
\begin{equation}
 \label{eq:pert_e_0}
  \etchiz{\Phigs}
  =  \frac{\elemm{\Chi_0}{\htcz + \lambda \vtc }{\Phigs}}{\braket{\Chi_0}{\Phigs}}.
\end{equation}
We now expand the ground-state energy in powers of $\lambda$
\begin{equation}
 \label{eq:pert_e_0_bis}
  \etchiz{\Phigs}
  =  \sum_{k=0}^\infty \lambda^k \tilde{E}^{(k)},
\end{equation}
which therefore implies to also expand the right eigenvector in powers of $\lambda$
\begin{equation}
 \ket{\Phigs}
 = \sum_{k=0}^\infty \lambda^k \ket{\Phi^{(k)}},
\end{equation}
where $\ket{\Phi^{(k)}}$ is the correction to the right ground state wave function at the order $k$ 
which is expanded in the basis of the right eigenvectors of $\htcz$ assuming intermediate bi-orthonormalization
\begin{equation}
 \begin{aligned}
  \ket{\Phi^{(k)}} = \sum_{i} c_i^{(k)} \ket{\Phi_i},\\
  \braket{\Chi_0}{\Phi^{(k)}} = 0 \quad \text{if }k\ne 0,
 \end{aligned}
\end{equation}
which implies, because of the bi-orthonormality property, that ${\braket{\Chi_0}{\Phigs}}=1$.

Truncating up to second order Eq. \eqref{eq:pert_e_0_bis} leads to 
\begin{equation}
 \label{eq:def_e_right_1}
 \begin{aligned}
  \etchiz{\Phigs}
                = &  \tilde{E}^{(0)} + \lambda  \tilde{E}^{(1)} +  \lambda^2 \tilde{E}^{(2)},  \\
 \end{aligned}
\end{equation}
which then yields
\begin{equation}
 \label{eq:def_e_right_2}
 \begin{aligned}
 &  \tilde{E}^{(0)} = \elemm{\Chi_0}{\htcz}{\Phi^{(0)}}, \\
 &  \tilde{E}^{(1)} = \elemm{\Chi_0}{\vtc}{\Phi^{(0)}} , \\
 &  \tilde{E}^{(2)} = \elemm{\Chi_0}{\vtc}{\Phi^{(1)}} .
 \end{aligned}
\end{equation}
To obtain the equation for the perturbed wave function one replaces the expressions of both
$\ket{\Phigs}$ and $\etchiz{\Phigs}$ in Eq. \eqref{eq:pert_e_0} and for $\Phi^{(1)}$, which then leads to
\begin{equation}
 \label{eq:pert_phi1}
 \htcz \ket{\Phi^{(1)}} + \vtc \ket{\Phi^{(0)}} = \tilde{E}^{(0)} \ket{\Phi^{(1)}} + \tilde{E}^{(1)} \ket{\Phi^{(0)}}.
\end{equation}
By projecting Eq. \eqref{eq:pert_phi1} on a function $\ket{\Chi_i}$ one obtains the expression of the coefficient of the right function at first order
\begin{equation}
 \label{eq:phi_1}
 c_i^{(1)} = \frac{\elemm{\Chi_i}{\vtc}{\Phi^{(0)}}}{\tilde{E}^{(0)} - \epsilon_i},
\end{equation}
and therefore one can obtain the second order contribution to the energy
\begin{equation}
 \tilde{E}^{(2)} = \sum_{i=1}^N \tilde{E}^{(2)}_i,
\end{equation}
where $\tilde{E}^{(2)}_i $ is the contribution at second order to the energy of the function $\ket{\Phi_i}$
\begin{equation}
 \label{eq:def_pt2}
 \begin{aligned}
 \tilde{E}^{(2)}_i & = \elemm{\Chi_0}{\vtc}{\Phi_i}\,c_i^{(1)}  \\
                   & = \frac{\elemm{\Chi_0}{\vtc}{\Phi_i} \elemm{\Chi_i}{\vtc}{\Phi^{(0)}} }{\tilde{E}^{(0)} - \epsilon_i}.
 \end{aligned}
\end{equation}
With respect to the standard Hermitian case, one can notice here several differences in Eqs. \eqref{eq:phi_1} and \eqref{eq:def_pt2}:
i) The first-order coefficient $c_i^{(1)}$ is computed using $\elemm{\Chi_i}{\vtc}{\Phi^{(0)}}$, implying therefore the use of the left function $\Chi_i$ satisfying $\braket{\Chi_i}{\Phi_i}=1$. This is a consequence of the bi-orthonormal framework, as in the case of an orthonormal framework, the function $\Chi_i$ would be simply equal to $\Phi_i$.
ii) The computation of the energy implies in the general case, through $\elemm{\Chi}{\vtc}{\Phi_i}$,
the use of a left function $\Chi$, that we chose here to be $\ket{\Chi_0}$.

One could also expand in perturbation the left eigenvector $\ket{\Chi_{\text{gs}}}$, and evaluate the functional at the right function $\ket{\Phi_0}$.
In that case, one would obtain exactly the same expansion for the energy up to second-order.

\subsubsection{Selected CI algorithm in a bi-orthonormal framework}
\label{sec:tcsci}
In Ref.~\onlinecite{AmmSceGin-JCP-22} we investigated the various flavours of TC-SCI using an orthonormal framework.
Among the different choices of selection criteria tested, we found that the one based on the second-order contribution
to the energy using both the left  and right eigenvectors of $\htcz$ as zeroth order wave function
was the most efficient because it allows to improve both the left  and right eigenvectors.
We therefore follow a similar path here, using an \ac{EN} zeroth-order Hamiltonian, 
and present our bi-orthonormal TC-SCI (BiO-TC-SCI) algorithm
which, at an iteration $n$, can be summarized as follows.
\begin{enumerate}
 \item A given zeroth order set of right Slater determinants
 $\mathcal{P}_r^n=\{\ket{\Phi_\rmi},\,\rmi=1,\dots,N_{\text{det}} \}$ is known, and therefore its associated left zeroth order
 set is also known
 $\mathcal{P}_l^n=\{\ket{\Chi_\rmi},\,\rmi=1,\dots,N_{\text{det}} \}$.
 One obtains then the ground state left  and right eigenvectors of the TC-Hamiltonian within $\mathcal{P}_l^n$
 and $\mathcal{P}_r^n$
\begin{equation}
 \label{eq:diago}
 \begin{aligned}
 \htc^\dagger \ket{\Chi^{(0)}} = \tilde{E}^{(0)} \ket{\Chi^{(0)}}, \\
 \htc \ket{\Phi^{(0)}} = \tilde{E}^{(0)} \ket{\Phi^{(0)}}, \\
 \end{aligned}
\end{equation}
 with
  \begin{equation}
   \begin{aligned}
 &   \ket{\Chi^{(0)}} = \sum_{\rmi \,\in\,\mathcal{P}_l^n} d_{\rmi}^{(0)} \ket{\Chi_\rmi},\\
 &   \ket{\Phi^{(0)}} = \sum_{\rmi \,\in\,\mathcal{P}_r^n} c_{\rmi}^{(0)} \ket{\Phi_\rmi}. \\
   \end{aligned}
  \end{equation}
 \item For each Slater determinant $\ket{\Phi_{\rmi}}\notin \mathcal{P}_r^n$, estimate its importance thanks to its
 contribution to the energy at second-order using the \ac{EN} zeroth-order Hamiltonian
 \begin{equation}
 \label{eq:pt2_sci}
 \tilde{E}^{(2)}_\rmi  =
 \frac{\elemm{\Chi^{(0)}}{\htc}{\Phi_\rmi} \elemm{\Chi_\rmi}{\htc}{\Phi^{(0)}} }{\tilde{E}^{(0)} - \epsilon_\rmi},
 \end{equation}
 where $\epsilon_\rmi = \elemm{\Chi_\rmi}{\htc}{\Phi_\rmi}$.
 We also compute on the fly the total second-order contribution to the energy
 \begin{equation}
 \label{eq:pt2_sci_tot}
 \tilde{E}^{(2)} = \sum_{\rmi \notin \mathcal{P}_r^n }\tilde{E}^{(2)}_\rmi,
 \end{equation}
and estimate the energy as
\begin{equation}
 \label{eq:e_tot_sci}
 \tccipsi = \tilde{E}^{(0)} + \tilde{E}^{(2)}.
\end{equation}
 \item Select the set of $N_{\Phi_{\rmi}}$ right Slater determinants $\left\{\ket{\Phi_{\rmi}}\right\}$, labelled $\mathcal{A}_r^n$,
 with the largest $\left|\tilde{E}^{(2)}_\rmi\right|$. This automatically defines the corresponding set of
 left Slater determinants $\mathcal{A}_l^n$.
 \item Add the set $\mathcal{A}_r^n$ to $\mathcal{P}_r^n$ and $\mathcal{A}_l^n$ to $\mathcal{P}_l^n$ to
   define the new set of both left  and right Slater determinants of the zeroth order space
  \begin{equation}
   \begin{aligned}
   \mathcal{P}_r^{n+1} = \mathcal{P}_r^n \cup \mathcal{A}_r^n,\\
   \mathcal{P}_l^{n+1} = \mathcal{P}_l^n \cup \mathcal{A}_l^n.\\
   \end{aligned}
  \end{equation}
 \item Go back to step~1 and iterate until a given convergence criterion is reached.
\end{enumerate}

\subsubsection{Technical details about the BiO-TC-SCI algorithm}
\label{sec:technique}
The computation of \ac{TC} Hamiltonian matrix elements for the diagonalization step of Eq.\eqref{eq:diago}
is done within the so-called 5-idx approximation introduced in Ref.~\onlinecite{DobCohAlaGin-JCP-22}
which consists in neglecting the pure triple excitation terms in the three-body operator,
\textit{i.e.} the terms involving integrals $\lmuijmkln$ with six distinct indices in Eq.\eqref{eq:def_hub_sec_q}.
In addition, when selecting a given Slater determinant we automatically include all other Slater determinants belonging to the configuration space functions (CSF)
in which the determinant is involved, such that pure spin states are obtained after diagonalization.
The diagonalization is made using two distinct Davidson procedures: one for the left and one for the right eigenvectors, following Ref.~\onlinecite{HirNak-JComP-82}.

As the computation of $\tilde{E}^{(2)}$ is costly,
we neglect all contributions from the three-electron operator in Eqs. \eqref{eq:pt2_sci} and \eqref{eq:e_tot_sci},
and also adapt the stochastic version proposed in Ref.~\onlinecite{GarSceLooCaf-JCP-17}.
In the original algorithm, external determinants are organized in batches generated from
a determinant of the internal space. The batch generated by $\ket{I}$ is drawn with a probability
$|c_I^{(0)}|^2 / \sum_J |c_J^{(0)}|^2$.
In the present work, the batches are drawn with a probability 
$|d_I^{(0)}  c_I^{(0)}| / \sum_J |d_J^{(0)}  c_J^{(0)}| $.

Finally, we use the extrapolation technique of Ref.~\onlinecite{HolUmrSha-JCP-17} which allows
to estimate the \ac{TC}-\ac{FCI} energy as the zeroth-order energy $\tilde{E}^{(0)}$ obtained for a vanishing $\tilde{E}^{(2)}$.

\subsection{A simple three-body correlation factor for frozen-core calculations}
\label{sec:three_e_j}
Although the BiO-TC-SCI algorithm presented here is applicable to any form of correlation factor,
we focus on a linearized version of the simple three-body correlation factor
developed in Ref.~\onlinecite{AmmSceGin-JCTC-2023}. The correlation factor reads
\begin{equation}
 \label{def:unew}
 \unew{1}{2} = \umu{1}{2} \envlop{1} \envlop{2},
\end{equation}
where $\umu{1}{2}$ is the one-parameter correlation factor introduced in Ref.~\onlinecite{Gin-JCP-21},
\begin{equation}
 \label{eq:def_j}
 \umu{1}{2} = \frac{1}{2}r_{12}\,\bigg( 1 - \text{erf}(\mu\, r_{12}) \bigg) - \frac{1}{2\sqrt{\pi}\mu}e^{-(\mu\, r_{12})^2},
\end{equation}
and the envelope
\begin{equation}
 \label{def:envlop}
 \envlop{}= 1 - \sum_{m=1}^{\nuclnum} \exp\big( - \alpha_m \big| \br{}-\brbig{m} \big|^2 \big),
\end{equation}
is the linearized version of the function introduced in Ref.~\onlinecite{AmmSceGin-JCTC-2023}, with $\brbig{m}$ the position of the $m$-th nucleus in the system.
The envelope $\envlop{}$ plays the role of a damping function which suppresses the effect of the correlation factor
$\umu{1}{2}$ near each nucleus.
The parameters $\alpha_m$ control the typical range on which the correlation factor $\umu{1}{2}$ is killed
by the envelope $\envlop{1}\envlop{2}$ around the nucleus located at $\brbig{m}$.
The advantages of this relatively simple correlation factor are that, as shown in Ref.~\onlinecite{AmmSceGin-JCTC-2023},
i) an efficient analytical-numerical scheme can be used to obtain the integrals $\kijkl$ and $\lmuijmkln$,
ii) provided that a typical valence value of the $\mu$ parameter is given ($\mu=0.87$), the correlation factor
has only one parameter for each nucleus in the molecule,
iii) the parameters $\alpha_m$ are transferable, and can be optimized only for the isolated atoms.

\section{Results}
\label{sec:results}
\subsection{Computational details}
The BiO-TC-SCI code, with all required integrals, has been implemented in the Quantum Package software~\cite{QP2-JCTC-19}. 
The computation of integrals is based on a mixed analytical-numerical integration scheme, as detailed in Ref.~\onlinecite{AmmSceGin-JCTC-2023}.

All computations are carried out using the cc-pVXZ Dunning family of basis
sets. The \ac{TC} calculations have been performed with the correlation factor as
detailed in Sec. \ref{sec:three_e_j}. For each atom with a nuclear charge
$Z_m$, the corresponding nuclear parameter $\alpha_m$ is determined as the
minimum of a \ac{VMC} calculation for that atom using a single Slater
determinant. This determinant is constructed with orbitals that are the right
eigenvalues of the TC-Fock operator in the cc-pVTZ basis set (for more details,
see Ref.~\onlinecite{AmmSceGin-JCTC-2023}).
The only exception is the hydrogen atom, for which we take $\alpha_{\text{H}}=\infty $, as it has no core electrons. We adhere to the strategy of having a unique nuclear parameter $\alpha_m$ for each atom, regardless of the basis set used, and we do not re-optimize these parameters in the molecular systems. This approach results in a correlation factor that does not require any system-specific optimization.

The parameters $\alpha_m$ used in this study are presented in Table \ref{tab:alpha_opt}. All the BiO-TC-SCI energies are calculated using the extrapolation scheme described in Sec. ~\ref{sec:technique}. Unless explicitly stated otherwise, all calculations are performed within the frozen core approximation with a [He] core.
\begin{table}[h]
\caption{\label{tab:alpha_opt}Parameters $\alpha_m$ used in the present work.}
\begin{ruledtabular}
\begin{tabular}{l|ccccccc}
 Atom       & H        &  Li &   C &   N &   O &   F &  Ne \\
 $\alpha_m$ & $\infty$ & 5.5 & 2.5 & 2.5 & 2.0 & 2.0 & 2.0 \\
\end{tabular}
\end{ruledtabular}
\end{table}

\subsection{Benefits of \BiO{} orbital optimization: freezing core orbitals and linear extrapolation}
\label{sec:benefit_bio}
Core electrons have an important property: they contribute minimally to most
chemically relevant energy differences, such as \acp{IP} or \acp{AE}. This
property is leveraged in virtually all post-\ac{HF} methods by using
the \ac{fc} approximation, significantly reducing the computational
cost, especially when using strong scaling methods like \ac{SCI} or \ac{CC}.
However, this technique depends on the optimization of core orbitals such that i) excitations of electrons from these orbitals to valence orbitals have a negligible weight in the wave functions of the low energy part of the spectrum, and ii) the core-core and core-valence correlation energy are essentially transferable from atoms to molecules.
The most straightforward way to achieve this decoupling between core and valence orbitals is to obtain eigenvectors of the Fock matrix, i.e., perform a canonical \ac{HF} optimization.

Transitioning to the context of the \ac{TC} Hamiltonian, given that the
effective interaction is no longer the Coulomb interaction, there is no reason
for the usual core electrons \ac{HF} orbitals to achieve the decoupling
necessary for the \ac{fc} approximation. A logical approach is then to perform
the equivalent of the \ac{HF} orbital optimization adapted for the \ac{TC}
Hamiltonian. This results in a bi-variational and \BiO{} framework as described
in Sec.~\ref{sec:orb_opt}.

To examine the impact of the quality of core orbitals used in \ac{TC}
calculations on typical valence energy differences, we computed the \acp{IP}
of oxygen, fluorine, and neon atoms, along with the \ac{AE}
of the F$_2$ molecule. These computations were performed in the
cc-pVTZ basis set, with or without the \ac{fc} approximation, using
conventional \ac{SCI}, or the \ac{TC}-\ac{SCI} with either \ac{RHF} orbitals or
\BiO{} orbitals. The results of these computations are presented in
Table~\ref{tab:ip}.
As can be seen from Table \ref{tab:ip}, the \ac{fc} approximation affects both
the \acp{IP} and \ac{AE} by a few tenths of mH when using the standard \ac{SCI}. This is
typically what is expected for such an energy difference driven by valence
properties.
In the context of the \ac{TC}-\ac{SCI} framework, a variation of the same order
of magnitude is observed when using the \BiO{} orbitals, 
while variation by an order of magnitude larger is observed when using the \ac{RHF} orbitals.
More importantly, it can be observed that when using \ac{RHF} orbitals, the
variation on the \acp{IP} with the \ac{fc} approximation increases with the nuclear charge. It is 1~kcal/mol,
1.4~kcal/mol, and 2.0~kcal/mol for the oxygen, fluorine, and neon atoms,
respectively. Also, the impact of the \ac{fc} approximation on the \ac{AE} of the
F$_2$ molecule is about 1.4~kcal/mol.
However, the all-electron \ac{TC} calculations using both \ac{RHF} and \BiO{}
orbitals agree within a few tenths of mH. The small deviation of the \ac{AE} from
the \ac{RHF} and \BiO{} calculations originates from the 5-idx approximation, which
does not guarantee strict orbital invariance as part of the Hamiltonian is
truncated.

\begin{figure}
        \centering
                \includegraphics[width=\linewidth]{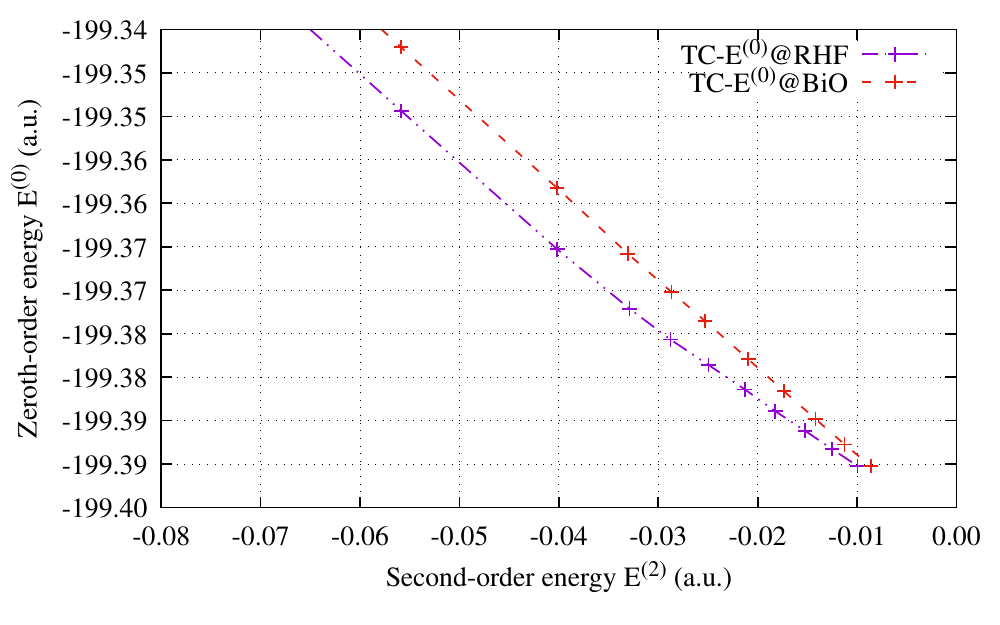}
                \includegraphics[width=\linewidth]{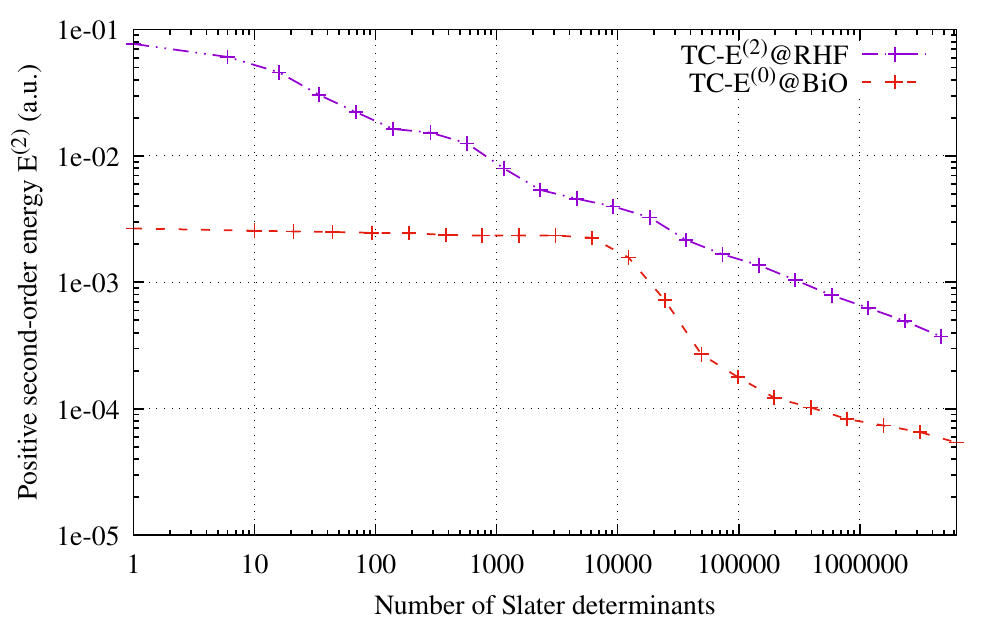}
        \caption{\label{extrap_f2_all_e}
 F$_2$, equilibrium geometry, cc-pVTZ basis set: a) Convergence of $\tilde{E}^{(0)}$ using either RHF orbitals
 (TC-$\text{E}^{(0)}$@RHF) or TC-SCF orbitals (TC-$\text{E}^{(0)}$@BiO)  as a function of $\tilde{E}^{(2)}$.
 b) Convergence of the positive contributions of $\tilde{E}^{(2)}$ using either TC-SCF or RHF orbitals.
        }
\end{figure}

Another notable characteristic of the \BiO{} orbitals is that they facilitate a
more straightforward linear extrapolation of $\tilde{E}^{(0)}$ as a function of
$\tilde{E}^{(2)}$. The latter is crucial for obtaining reliable estimates of the
\ac{TC}-\ac{FCI} energy. 
To illustrate this, we present in Fig.~\ref{extrap_f2_all_e} the variation of
$\tilde{E}^{(0)}$ as a function of $\tilde{E}^{(2)}$ for the all-electron
calculation of the F$_2$ molecule in the cc-pVTZ basis set. As can be seen from
Fig.~\ref{extrap_f2_all_e}, the \ac{TC}-\ac{SCI} exhibits more linearity when performed
using \BiO{} orbitals than when using \ac{RHF} orbitals.
Upon careful examination of the data, it appears that this difference arises
from the amount of positive contribution to $\tilde{E}^{(2)}$, which is
typically ten times larger when using \ac{RHF} orbitals, as illustrated in
Fig.~\ref{extrap_f2_all_e}.

These results underscore the importance of optimizing the orbitals within a
\BiO{} framework when performing \ac{TC} calculations. This optimization is key for enabling the \ac{fc}
approximation along with extrapolation techniques, which are essential for the
application of the method to large systems and/or active spaces.

\begin{table}[h]
\caption{\label{tab:ip}Ionization potentials (IP) and atomization energy (AE), in mH, computed with TC-SCI in a cc-pVTZ basis set, with (fc) or without (all-e) the frozen core approximation, and using either the usual HF orbitals (RHF) 
orbitals or the \BiO{} \ac{TC}-SCF orbitals (BiO). Usual hermitian SCI values are also reported.}
\begin{ruledtabular}
\begin{tabular}{l c c c}
         &  SCI  &   TC-SCI@RHF & TC-SCI@BiO \\
\hline
         & \multicolumn{3}{c}{IP of the oxygen atom} \\
all-e    & 489.85 &  496.04 &  495.87 \\
fc       & 489.75 &  497.74 &  495.86 \\
\hline
         & \multicolumn{3}{c}{IP of the fluorine atom} \\
all-e    & 630.00 &  636.82 &  636.62 \\
fc       & 629.80 &  639.08 &  636.55 \\
\hline
         & \multicolumn{3}{c}{IP of the neon atom} \\
all-e    & 783.09 &  789.85 &  789.88 \\
fc       & 782.80 &  793.02 &  789.68 \\
\hline
         & \multicolumn{3}{c}{AE of the F$_2$ molecule} \\
all-e    & 56.76  &  61.05  & 61.69   \\
fc       & 56.39  &  63.28  & 61.47   \\
\end{tabular}
\end{ruledtabular}
\end{table}

\subsection{Convergence of the BiO-TC-SCI algorithm}
\label{sec:conv_tc_cipsi}

\begin{figure}
        \centering
                \includegraphics[width=\linewidth]{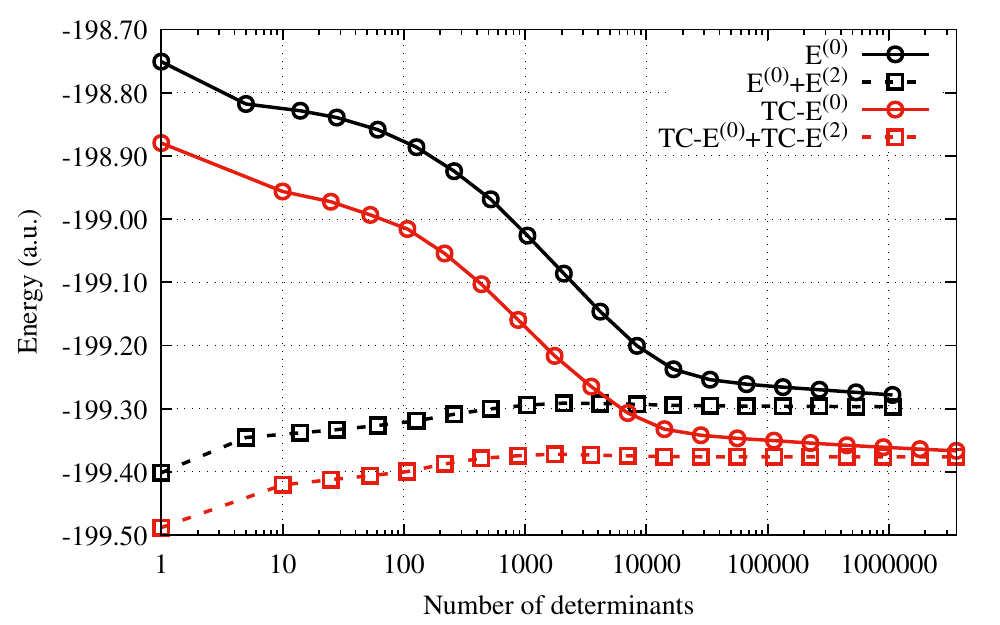}
                \includegraphics[width=\linewidth]{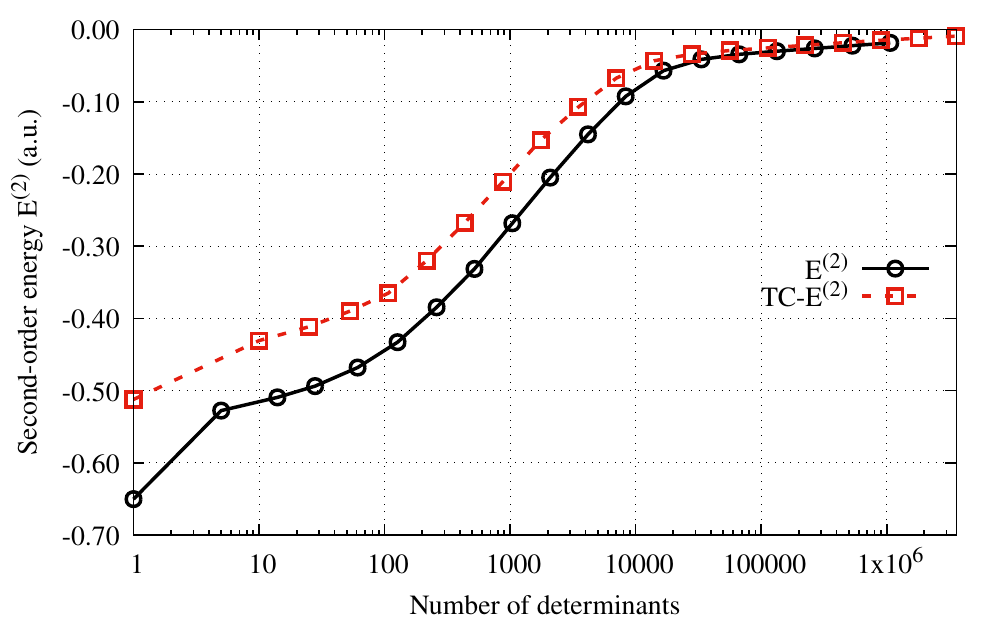}
                \includegraphics[width=\linewidth]{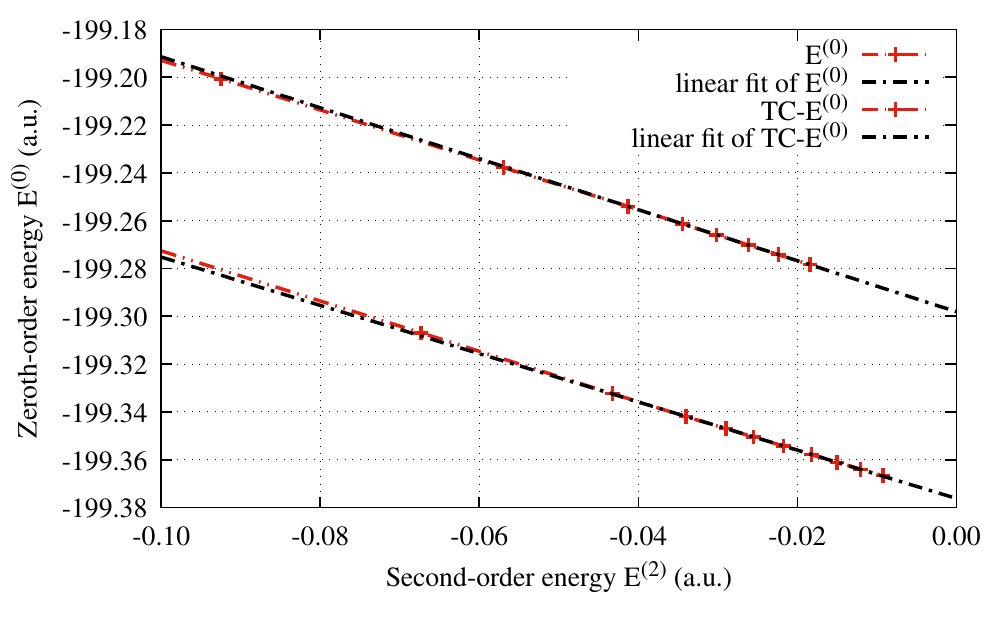} \\
        \caption{\label{fig:conv_f2}
 F$_2$, equilibrium geometry, cc-pVTZ basis set: 
 Convergence of $\tilde{E}^{(0)}$ (TC-$E^{(0)}$), $\tilde{E}^{(2)}$ (TC-$E^{(2)}$), and $\tccipsi$ (TC-$E^{(0)}$+TC-$E^{(2)}$)
 (see Eqs. \eqref{eq:diago}, \eqref{eq:pt2_sci_tot}, and \eqref{eq:e_tot_sci}, respectively) 
 as a function of the number
 of Slater determinants. The corresponding usual quantities are also reported for comparison,
 and are denoted without the ``TC'' prefix. The convergence of $\tilde{E}^{(0)}$ as a function of $\tilde{E}^{(2)}$ is also reported, together with a linear fit of the data.
        }
\end{figure}

\begin{figure}
        \centering
                \includegraphics[width=\linewidth]{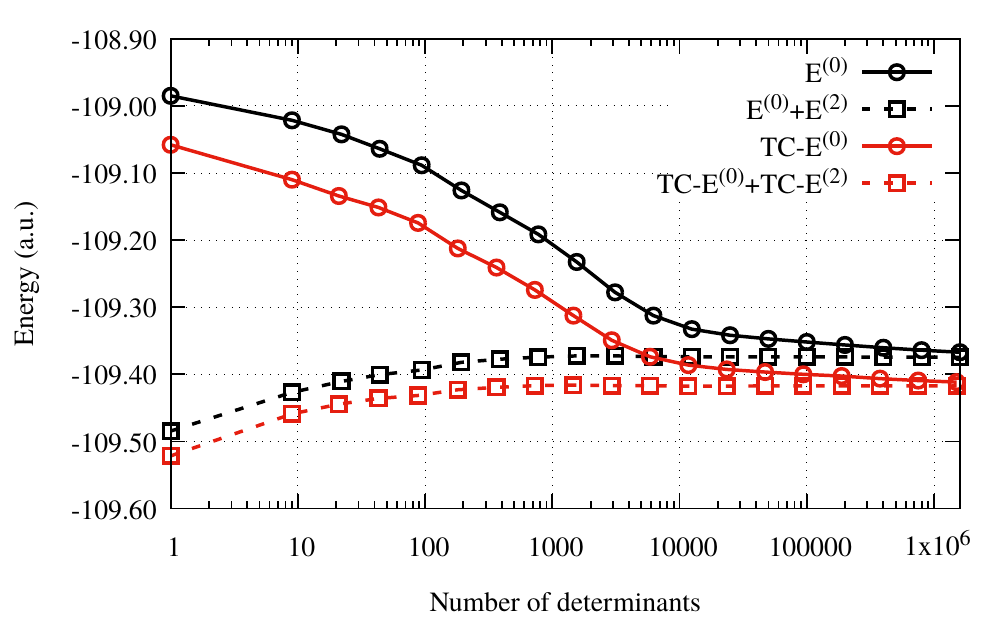}
                \includegraphics[width=\linewidth]{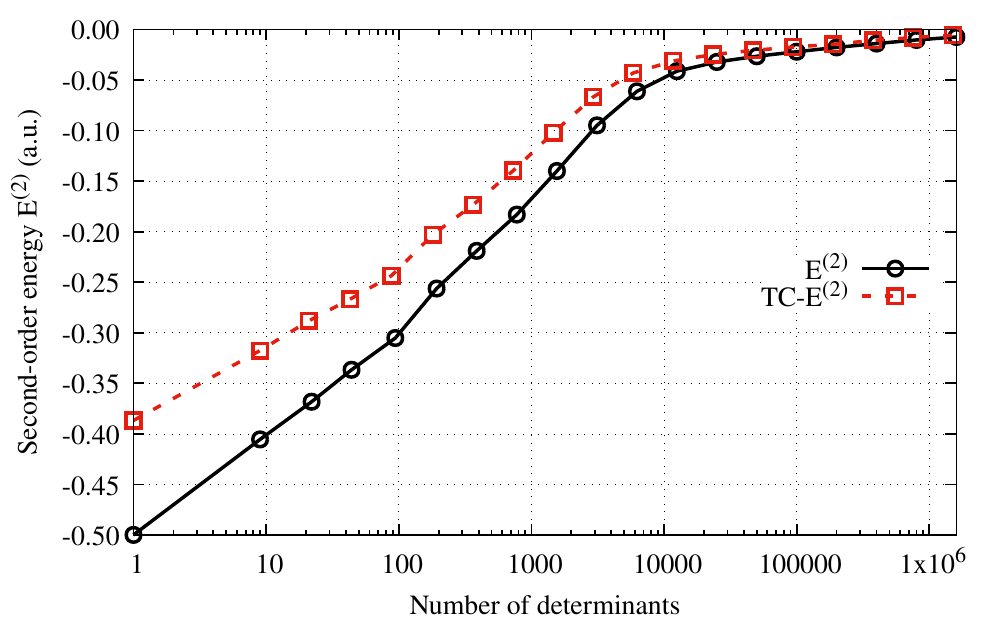}
                \includegraphics[width=\linewidth]{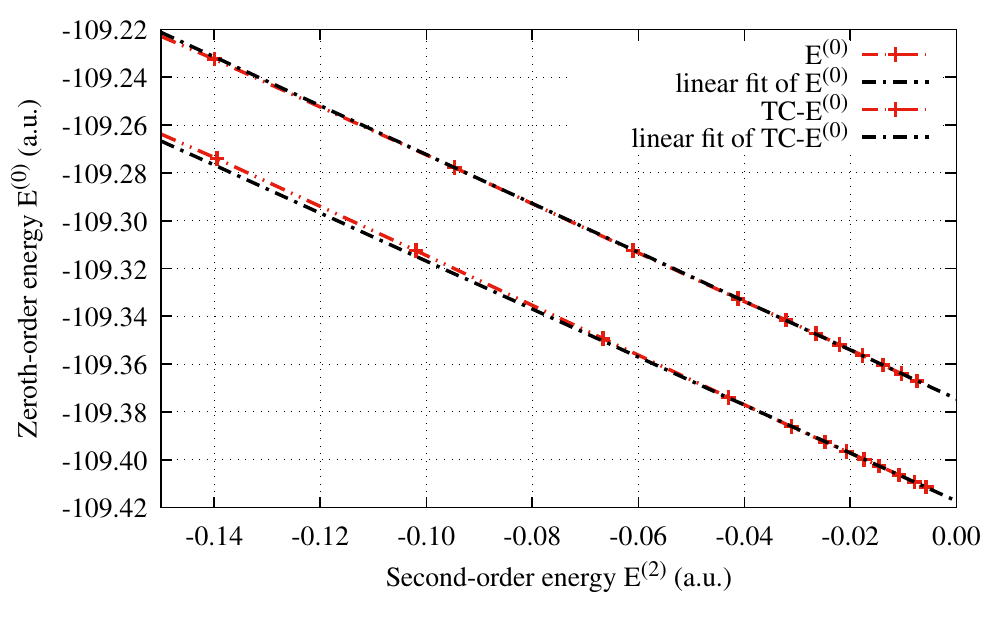} \\
        \caption{\label{fig:conv_n2}
 N$_2$, equilibrium geometry, cc-pVTZ basis set:
 Convergence of $\tilde{E}^{(0)}$ (TC-$E^{(0)}$), $\tilde{E}^{(2)}$ (TC-$E^{(2)}$), and $\tccipsi$ (TC-$E^{(0)}$+TC-$E^{(2)}$)
 (see Eqs. \eqref{eq:diago}, \eqref{eq:pt2_sci_tot}, and \eqref{eq:e_tot_sci}, respectively) 
 as a function of the number
 of Slater determinants. The corresponding usual quantities are also reported for comparison,
 and are denoted without the ``TC'' prefix. The convergence of $\tilde{E}^{(0)}$ as a function of $\tilde{E}^{(2)}$ is also reported, together with a linear fit of the data.
        }
\end{figure}

\begin{figure}
        \centering
                \includegraphics[width=\linewidth]{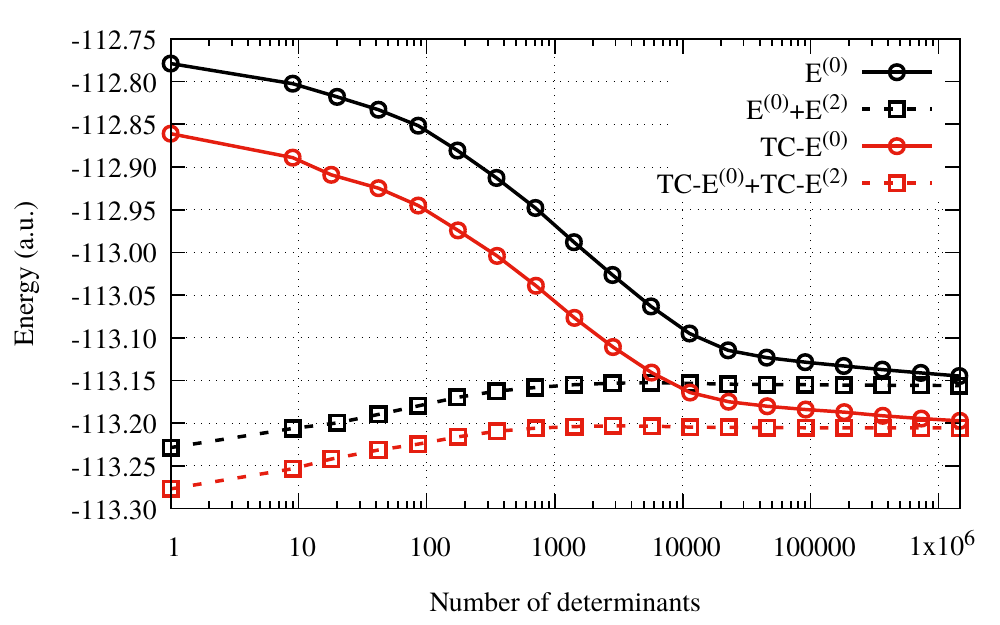}
                \includegraphics[width=\linewidth]{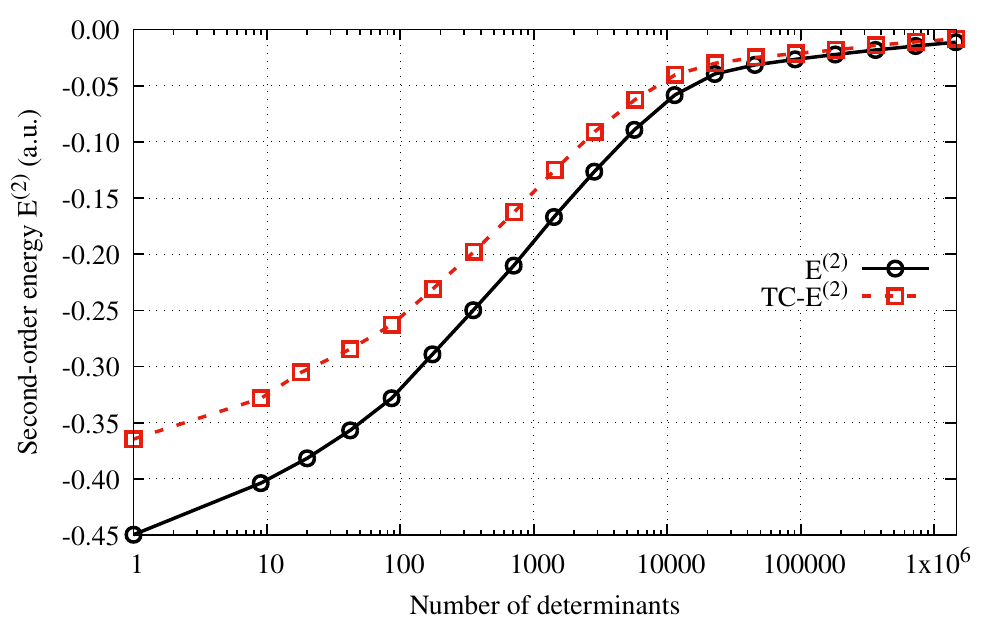}
                \includegraphics[width=\linewidth]{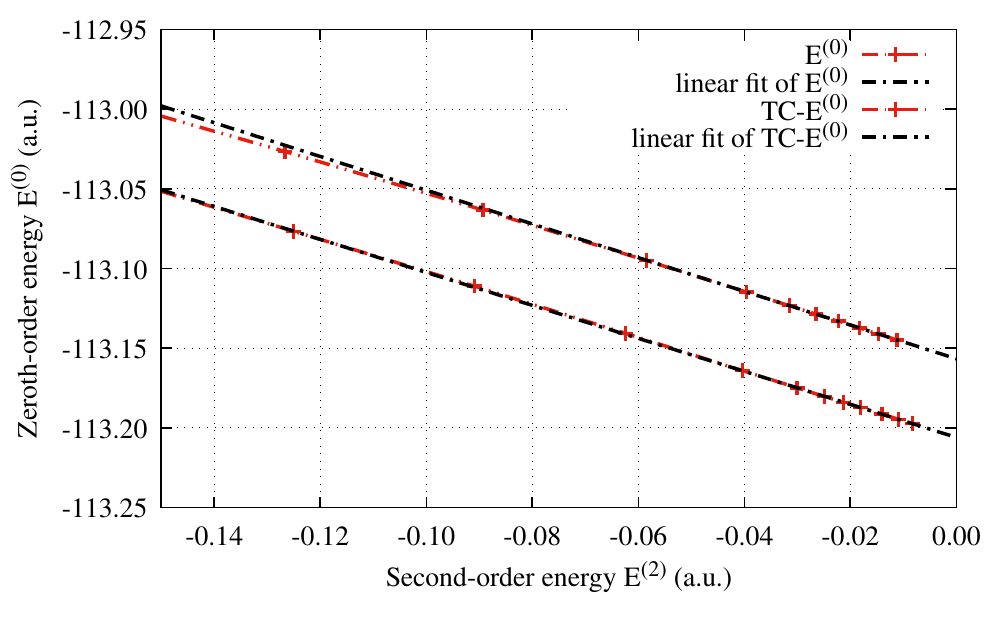} \\
        \caption{\label{fig:conv_co}
 CO, equilibrium geometry, cc-pVTZ basis set: 
 Convergence of $\tilde{E}^{(0)}$ (TC-$E^{(0)}$), $\tilde{E}^{(2)}$ (TC-$E^{(2)}$), and $\tccipsi$ (TC-$E^{(0)}$+TC-$E^{(2)}$)
 (see Eqs. \eqref{eq:diago}, \eqref{eq:pt2_sci_tot}, and \eqref{eq:e_tot_sci}, respectively) 
 as a function of the number
 of Slater determinants. The corresponding usual quantities are also reported for comparison,
 and are denoted without the ``TC'' prefix. The convergence of $\tilde{E}^{(0)}$ as a function of $\tilde{E}^{(2)}$ is also reported, together with a linear fit of the data.
        }
\end{figure}

To illustrate the convergence of the current BiO-TC-SCI algorithm, we present
the convergence of $\tilde{E}^{(0)}$, $\tilde{E}^{(2)}$ and $\tccipsi$ (see
Eqs. \eqref{eq:diago}, \eqref{eq:pt2_sci_tot}, and \eqref{eq:e_tot_sci},
respectively) as a function of the number of Slater determinants for the
$\text{F}_2$, $\text{N}_2$, and $\text{CO}$ molecules in the cc-pVTZ basis set
(Figs. \ref{fig:conv_f2}, \ref{fig:conv_n2}, and \ref{fig:conv_co},
respectively). We also include the extrapolation towards the TC-FCI energy, as
is typically done in SCI calculations. For comparison, the convergence of the
standard SCI scheme is also reported.

From Figs.~\ref{fig:conv_f2}, \ref{fig:conv_n2} and \ref{fig:conv_co}, it clearly appears that the convergence of the BiO-TC-SCI algorithm
is at least as fast as in usual \ac{SCI} algorithms, and that it can also be effectively extrapolated using a linear fitting.
When examining the convergence curve of $\tilde{E}^{(2)}$ as a function of the
number of Slater determinants, it is apparent that $|\tilde{E}^{(2)}|$ is
consistently smaller than when using the bare Hamiltonian. This is due to the
correlation factor already accounting for a portion of the correlation.
However, the correlation factor proposed in this study is not re optimised for each molecular situation, 
and therefore a significant part of the electron-electron correlation need to be represented
by the determinantal component of the wave function. This is why the typical
effect of wave function compaction is not significantly evident. Employing more
advanced forms of Jastrow factors would further reduce $|\tilde{E}^{(2)}|$,
leading to a faster convergence of the \ac{SCI} energy.

\subsection{Ability to break multiple covalent bonds and size consistency}
\label{sec:size_cons}

\begin{figure}
        \centering
                \includegraphics[width=\linewidth]{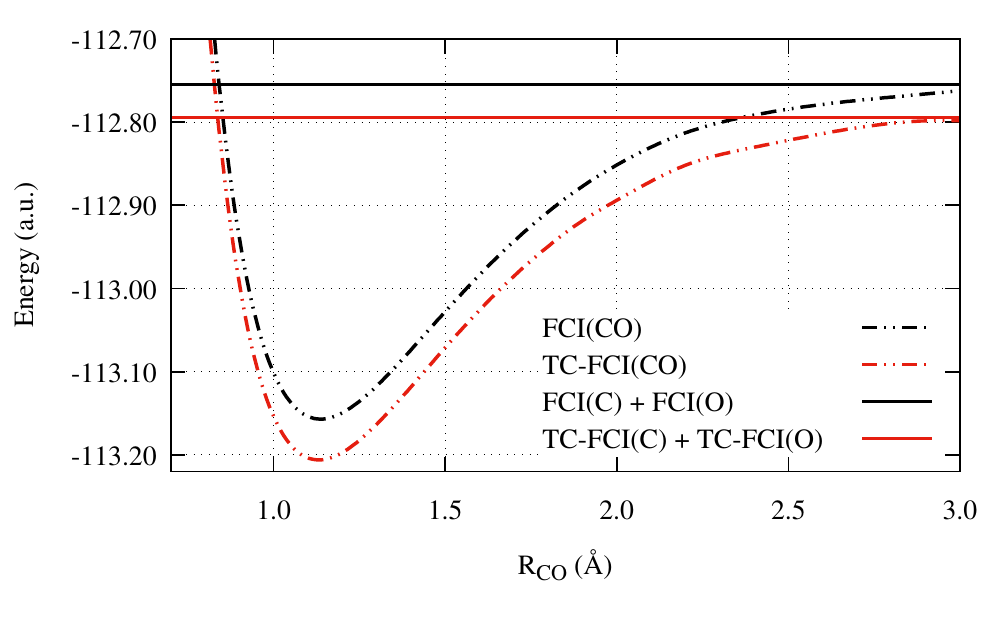}
        \caption{\label{fig:diss_co}
        CO, cc-pVTZ basis set: Potential energy curve computed using the extrapolated energies
	$\fci$ and $\tcfci$.
	The sum of the extrapolated $\fci(\text{C})+\fci(\text{O})$ and $\tcfci(\text{C})+\tcfci(\text{O})$ energies for
	the isolated atoms are also reported.
        }
\end{figure}

To explore the potential of the current BiO-TC-SCI approach in handling
strongly correlated systems, we present in Fig.~\ref{fig:diss_co} the potential
energy surface (PES) of the CO molecule, using the cc-pVTZ basis set, up to the
full dissociation limit. This is compared with the sum of the TC energies
obtained at the same level of calculation for the isolated atomic systems.
As can be seen from Fig.~\ref{fig:diss_co}, the PES is smooth across the curve
and converges towards the correct asymptotic limit, even for a non-homogeneous
system. This is because the extrapolated BiO-TC-SCI achieves near TC-FCI
quality, and the three-body correlation factor used in this study is size
consistent.

Additionally, we incorporate the PES obtained through the standard extrapolated FCI
into the same figure. A comparison between the two PESs reveals a more
pronounced and deeper well in the curve for the BiO-TC-SCI approach. This
result further substantiates the enhanced ability of BiO-TC-SCI to effectively
manage the electron correlation effects within the system.

\subsection{Atomization energies}
\label{sec:atomization}

\begin{figure}
        \centering
	\includegraphics[width=\linewidth]{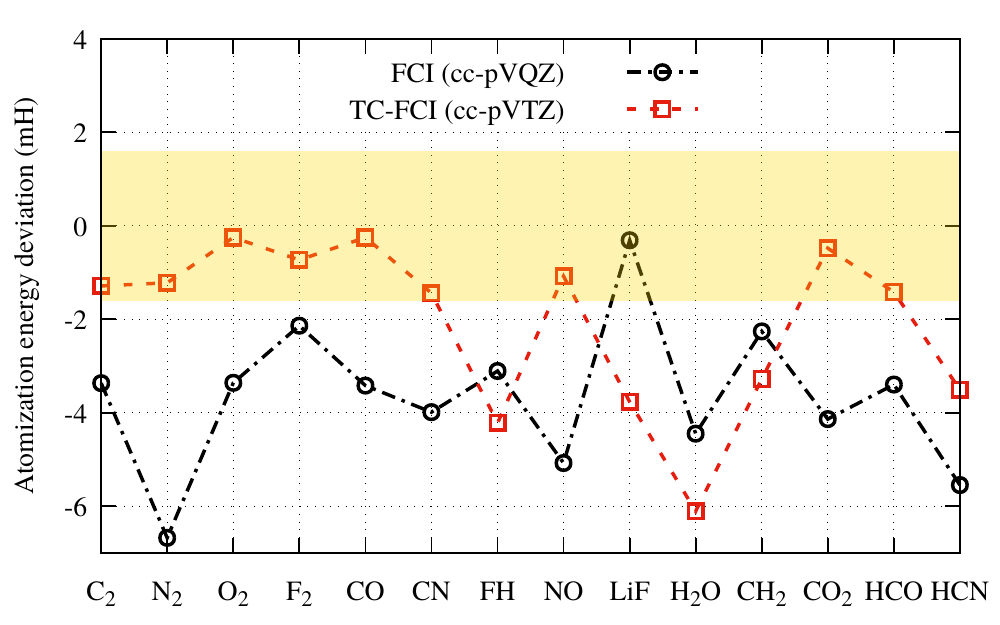}
        \caption{\label{fig:atom_ener}
	Deviation of $\fci$ and $\tcfci$ from the estimated exact atomization energies
	(in mH) in cc-pVQZ and cc-pVTZ, respectively.
	The deviation of each method from the reference iis defined using the convention
	$\delta E_{\text{method}} = E_{\text{method}} - E_{\text{estimated exact}}$.
	The figure includes a highlighted filled region, indicating the domain of chemical accuracy.
	Refer to Table~\ref{mol_table_deltae} for the specific values.
        }
\end{figure}

Table~\ref{mol_table_deltae} presents the atomization energy results (in mH)
for a selection of molecules using the standard extrapolated FCI or the extrapolated TC-FCI.
These calculations were performed in various basis sets, specifically cc-pVDZ,
cc-pVTZ, and cc-pVQZ for the FCI, and cc-pVDZ and cc-pVTZ for the TC-FCI. 
In addition to the calculated results, the ``estimated exact'' column provides an
estimation of the exact non-relativistic atomization energies, sourced from
Ref.~\onlinecite{YaoGinLiTouUmr-JCP-20}. This serves as a benchmark for the
accuracy of the methods used. The molecular geometries were also taken from
Ref.~\onlinecite{YaoGinLiTouUmr-JCP-20}.
To supplement the data in the table, we provide a visual representation of the
FCI ($\fci$) and TC-FCI ($\tcfci$) results in the cc-pVQZ and cc-pVTZ basis sets,
respectively, in Figure \ref{fig:atom_ener}. This aids in interpreting and
demonstrating the observed trends.

The results reveal the significant benefits of the TC-FCI calculation in terms of
convergence and accuracy across a range of molecules. For instance, for the
CO$_2$ molecule, the TC-FCI delivers atomization energies of 622.86~mH and
618.14~mH in the cc-pVDZ and cc-pVTZ basis sets, respectively. These values
indicate a convergence towards highly accurate results, closely approximating
the estimated exact value of 618.62~mH.
Similarly, for O$_2$, the TC-FCI provides atomization energies of 191.84~mH and
191.75~mH in the cc-pVDZ and cc-pVTZ basis sets, respectively. These values
show remarkable alignment with the estimated exact value of 192.0~mH, further
emphasizing the accuracy and convergence of the TC-FCI.
In contrast, the conventional FCI calculation, even with the larger cc-pVQZ basis
set, has difficulty converging to chemical accuracy for many molecules. This is
evident in the results observed for various systems, such as C$_2$, N$_2$, and
HCO, where the TC-FCI consistently surpasses the FCI in terms of accuracy and
convergence, particularly in the modest cc-pVTZ basis set.

The TC-FCI/cc-pVTZ consistently outperforms the FCI/cc-pVQZ.  However, in a few
cases such as LiF (with 216.39~mH in the TC-FCI/cc-pVTZ versus 219.85~mH in
the FCI/cc-pVQZ), the TC-FCI/cc-pVTZ does not exceed the efficiency of
the FCI/cc-pVQZ. This observation could be ascribed to the use of a relatively
simple correlation factor. Our future work will aim to explore the application
of the TC-FCI with more sophisticated Jastrow factors to further enhance its
performance and accuracy.

\begin{table*}[t]
\caption{\label{mol_table_deltae}
	Atomization energies (in mH) obtained using the standard extrapolated FCI or the extrapolated TC-FCI in different basis sets.
	}
\begin{ruledtabular}
\begin{tabular}{lcccccc}
\multirow{2}{*}{} & \multicolumn{3}{c}{CIPSI} & \multicolumn{2}{c}{TC-CIPSI} & \multirow{2}{*}{Estimated exact~\cite{YaoGinLiTouUmr-JCP-20}} \\
                  & cc-pVDZ & cc-pVTZ & cc-pVQZ & cc-pVDZ & cc-pVTZ &    \\
\hline
%
C$_2$             & 207.50 & 223.11 & 228.83 & 226.03 & 230.91 & 232.2 \\
%
N$_2$             & 319.13 & 345.72 & 356.02 & 359.67 & 361.48 & 362.7 \\
%
O$_2$             & 167.84 & 182.28 & 188.64 & 191.84 & 191.75 & 192.0 \\
%
F$_2$             & 44.38 & 56.39 & 60.06 & 58.18 & 61.47 & 62.2 \\
%
CO                & 385.95 & 401.50 & 408.68 & 413.52 & 411.85 & 412.1 \\
%
CN                & 254.40 & 274.40 & 283.21 & 282.24 & 285.76 & 287.2 \\
%
FH                & 201.64 &  218.31 & 223.07 & 217.59 & 221.96 & 226.18 \\
%
NO                & 211.60 & 229.88 & 237.67 & 242.32 & 241.67 & 242.75 \\
%
LiF               & 196.06 & 211.84 & 219.85 & 210.15 & 216.39 & 220.16 \\
%
H$_2$O            & 333.80 & 359.09 & 367.19 & 355.33 & 365.53 & 371.64 \\
%
CH$_2$\footnotesize{$^a$} & 263.80 & 281.80 & 286.34 & 275.75 & 285.32 & 288.6 \\
%
CO$_2$            & 569.29 & 604.20 & 614.48 & 622.86 & 618.14 & 618.62 \\
%
HCO               & 407.83 &  430.85 & 439.78 & 438.92 & 441.76 & 443.18 \\
%
HCN               & 451.74 & 481.59 & 491.51 & 488.25 & 493.55 & 497.06 \\
\end{tabular}
\end{ruledtabular}
\begin{minipage}{\textwidth}
\vspace{0.20 cm}
\begin{flushleft}
	\footnotesize{$^a$ $^1 A_1$ state.} \\
\end{flushleft}
\end{minipage}
\end{table*}

\section{Conclusion}
\label{sec:conclu}

In this work, we introduced the theoretical framework of \ac{SCI}
using a set of \BiO{} orbitals within the context of \ac{TC}
calculations.  We employed a linearized version of a recently developed
inexpensive three-body correlation factor~\onlinecite{AmmSceGin-JCTC-2023}, which
eliminates the need for re-optimizing the correlation factor for each molecule
and facilitates an efficient analytical/numerical evaluation of the integrals
required in TC calculations.
After establishing the main equations, we explored various aspects of the current
approach numerically. We first examined the benefits of using \BiO{} orbitals in
\ac{TC} calculations by studying the impact of the \ac{fc} approximation on
a set of \acp{IP} and \acp{AE}. Our findings indicate that the \ac{fc} approximation
has a similar impact in \ac{TC} calculations using \BiO{} orbitals as in
standard wavefunction techniques (typically a few 0.1~mH), 
while an order of magnitude greater variation is observed when using the \ac{RHF} orbitals.
These
observations support the idea of conducting \ac{TC} calculations using a \BiO{}
framework, as the \ac{fc} approximation is essential for handling large systems.
We then investigated the convergence of the current BiO-TC-SCI framework
compared to standard \ac{SCI} approaches. Our results show that similar convergence
can be expected, and that linear extrapolation is feasible, making our
BiO-TC-SCI approach as robust as any standard \ac{SCI} approach. We also examined
the ability of the BiO-TC-SCI algorithm to handle strongly correlated systems
by breaking the CO double bond, and demonstrated that the size consistency
property is fulfilled.
Finally, we assessed the quality of the current correlation factor on a set of
14 small organic molecules. Our results show that even with such a simple
correlation factor, the accuracy is typically better than a standard quadruple
zeta calculation. It is important to note that thanks to the specific form of
the correlation factor used here, no optimization of the latter was needed for
molecular calculations.

\begin{acknowledgments}
This work was performed using HPC resources from GENCI-TGCC
(gen1738,gen12363) and from CALMIP (Toulouse) under allocation
P22001, and was also supported by the European Centre of
Excellence in Exascale Computing TREX --- Targeting Real Chemical
Accuracy at the Exascale. This project has received funding from the
European Union's Horizon 2020 --- Research and Innovation program ---
under grant agreement no.~952165. 
Emmanuel Giner would like to thank Julien Toulouse for fruitful discussions 
regarding the bi orthogonal basis sets. 
\end{acknowledgments}


\bibliography{paper_tc_cipsi}
 \end{document}